\documentclass[letterpaper,twocolumn,10pt]{article}
\usepackage{usenix}
\usepackage{graphicx}
\graphicspath{{Figures/}}
\DeclareGraphicsExtensions{.pdf,.png,.jpg,.jpeg}
\usepackage{algorithmic}
\usepackage{booktabs}
\usepackage{multirow}
\usepackage{amsmath}
\usepackage{amssymb}
\usepackage{array}
\usepackage{xspace}
\usepackage{enumitem}
\usepackage{dsfont}
\usepackage{xcolor}
\usepackage{caption}

\newcommand{\appalgref}[1]{\hyperref[#1]{Appendix Algorithm~\ref*{#1}}}

\newcommand{\mypara}[1]{\medskip\noindent\textbf{#1.} \xspace}

\newcommand{\rep}[1]{\textrm{rep-}#1}
\newcounter{ieeealgorithm}
\renewcommand{\theieeealgorithm}{\arabic{ieeealgorithm}}

\newenvironment{ieeealgorithm}[2][]{%
    \par
    \refstepcounter{ieeealgorithm}%
    \noindent
    \begin{minipage}{\columnwidth}
    \hrule height 0.8pt
    \vspace{0.6ex}
    \noindent\textbf{Algorithm~\theieeealgorithm: #2}\par
    \vspace{0.6ex}
    \hrule height 0.4pt
    \vspace{0.8ex}
 }{%
    \vspace{0.6ex}
    \hrule height 0.8pt
    \end{minipage}
    \par
}

\begin{document}

\date{}
\title{\Large \bfseries ChronoPatch: Black-Box Energy-Latency Attacks on Vision-Language Models}

\author{
{\rm Jiawei Xing}\\
Shandong University
\and
{\rm Li Wang}\\
Shandong University
\and
{\rm Xiangtao Meng}\\
Shandong University
\and
{\rm Chuanchao Zang}\\
Shandong University
\and
{\rm Rui Wen}\\
Institute of Science Tokyo
\and
{\rm Zheng Li}\\
Shandong University
} 

\maketitle

\begin{abstract}
Vision-language models (VLMs) are increasingly deployed as generative inference services, where autoregressive decoding makes response length a key determinant of latency and computational cost. This creates an availability risk: adversarial visual inputs can prolong generation and increase serving overhead. Existing energy-latency attacks typically rely on gradients, logits, or token-level feedback, leaving open whether such risks persist when an attacker observes only the final response. We present ChronoPatch, a black-box attack that induces prolonged VLM generation requiring only the final response. It discovers visual delay anchors through response-length-guided evolutionary search, then minimizes perturbation magnitude through subspace-calibrated decision-based optimization. Across four open-source VLMs, ChronoPatch increases inference latency by 47.86\%--293.34\% and GPU energy consumption by 41.44\%--249.70\% under a 5\% spatial budget. It also induces more repetitive outputs, as reflected by increased n-gram repetition ratios. On three commercial VLM APIs, ChronoPatch increases response length by 37.6\%–81.5\% and client-observed latency by 1.82$\times$--2.29$\times$ under query budgets of 500 per API. These results reveal a generation-based availability risk that persists under final-response-only access and motivate generation-aware safeguards for deployed generative AI services.
\end{abstract}

\section{Introduction}
\label{sec:introduction}

Large vision-language models (VLMs) are increasingly deployed as multimodal inference services, serving as the perception and reasoning components in embodied agents, autonomous perception pipelines, and interactive decision-making systems~\cite{li2022mplug,li2023blip,chen2310minigpt,bai2023qwen,zhang2023multimodal}. In these deployments, availability depends not only on response quality but also on predictable latency and computational cost. Delayed perception or excessively long responses can postpone downstream decisions and increase serving overhead.

This exposes a security-relevant risk in autoregressive VLMs~\cite{alayrac2022flamingo,liu2023visual,awadalla2023openflamingo}. Given an image and a textual prompt, a VLM generates its response token sequentially until an end-of-sequence (EOS) token is emitted or the generation budget is exhausted~\cite{vaswani2017attention,gao2024inducing}. Since each generated token incurs an additional decoding step, generation length is a key determinant of inference latency and energy consumption~\cite{kwon2023efficient,gao2024inducing}. Adversarial inputs that delay response termination can therefore amplify inference cost without modifying model parameters or textual prompts. We refer to this threat as an \textit{energy-latency attack}~\cite{shumailov2021sponge}.

\begin{figure}[!t]
\centering

\includegraphics[width=\columnwidth]{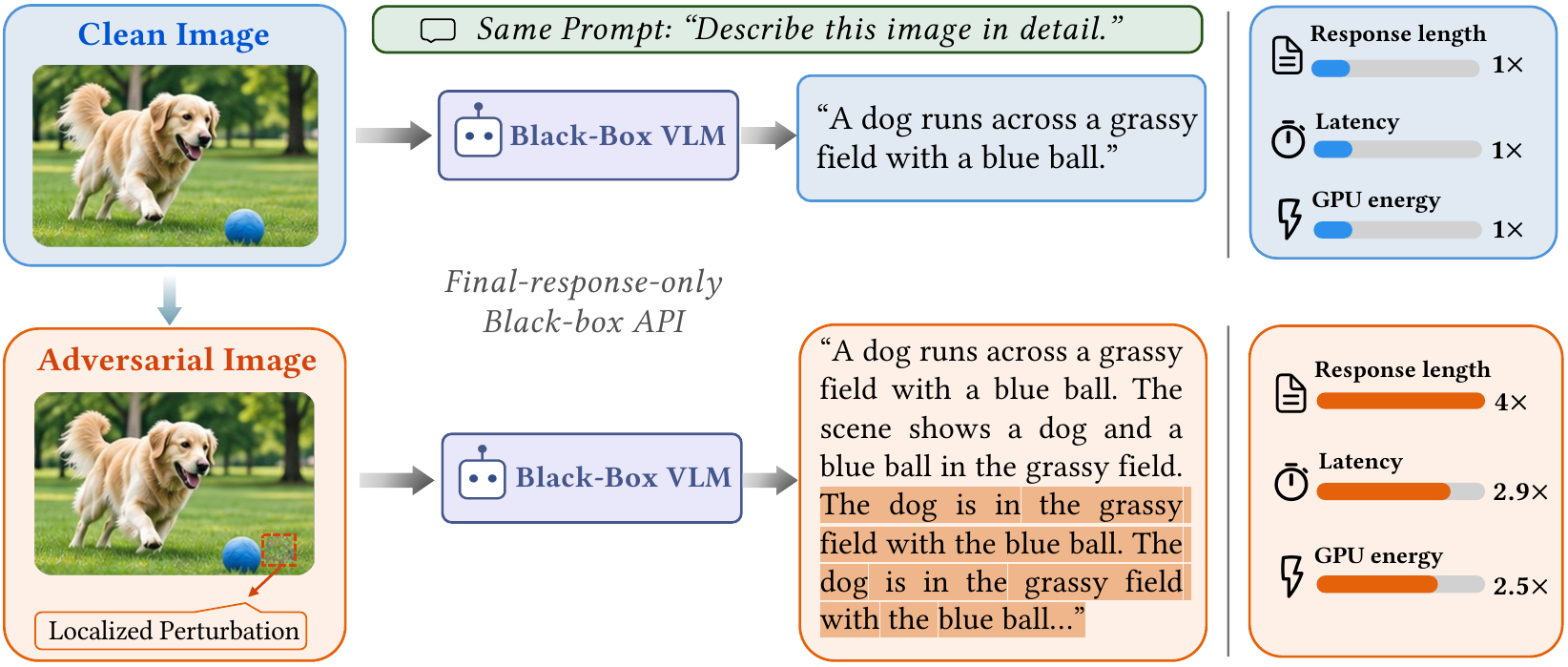}
\caption{Motivating example of ChronoPatch. Localized visual perturbations induce prolonged and repetitive VLM generation under final-response-only black-box access.}
\label{fig:chronopatch_demo}
\end{figure}

Prior energy-latency attacks demonstrate that adversarial inputs can amplify VLM inference costs, but differ in their feedback assumptions and attack surfaces. White-box methods such as NICGSlowDown~\cite{chen2022nicgslowdown} and Verbose Images~\cite{gao2024inducing} rely on model internals or gradients, whereas VLMInferSlow~\cite{wang2025vlminferslow} optimizes visual perturbations using continuous feedback such as output logits. Recent approaches also induce excessive reasoning through visual scene-text triggers~\cite{han2026overthink}, jointly optimize images and textual prompts~\cite{ni2026multimodal}, or exploit serving-side KV-cache and scheduling mechanisms~\cite{wang2026latencydos}. In contrast, ChronoPatch focuses on localized visual perturbations with a fixed textual prompt, using only completed-response length as optimization feedback, without gradients, logits, or intermediate decoding signals. This raises a central question: \textit{Can an attacker substantially amplify VLM generation cost using only final-response feedback?}

We study this final-response-only black-box setting and identify two challenges. First, no natural adversarial initialization is available to reliably induce prolonged generation. The attacker must discover effective visual delay patterns using only response-length feedback. Second, once a visual delay anchor is found, the attacker must reduce its perturbation magnitude while preserving prolonged generation, despite lacking intermediate feedback to guide refinement.

To address these challenges, we propose {ChronoPatch}, a two-stage energy-latency attack framework. Phase I, \textit{delay-anchor discovery}, uses response-length-guided evolutionary search over visual patterns and locations to identify visual delay anchors that induce prolonged generation. Phase II, \textit{perturbation refinement}, starts from these anchors and progressively reduces perturbation magnitude within the fixed editable region while preserving the long-generation effect. It employs subspace-calibrated decision-based optimization to adapt the exploration scale to the effective dimensionality of the editable region, enabling more reliable boundary estimation under spatial constraints.

We evaluate ChronoPatch on four representative open-source VLMs across three benchmark datasets, measuring generation length, inference latency, and GPU energy consumption. Under a perturbation budget of 5\% of the image area, ChronoPatch increases end-to-end inference latency by up to 293.34\% and GPU energy consumption by up to 249.70\% on MSCOCO. The generated responses also exhibit repetitive degeneration, particularly on Qwen3-VL and InternVL3. We further evaluate ChronoPatch on three commercial VLM APIs under a reduced query budget, observing response-length increases of 37.6\%--81.5\% and client-observed end-to-end latency reaching 1.82$\times$--2.29$\times$ that of clean requests, demonstrating that over-generation remains externally observable in deployed services.

Our contributions are summarized as follows:
\begin{itemize}
    \item We identify a final-response-only energy-latency threat in VLM services, showing that generation cost can be amplified using only returned response lengths, without access to model internals or intermediate feedback.

    \item We propose ChronoPatch, a two-stage attack framework that discovers visual delay anchors from response-length feedback and refines them into low-distortion perturbations under constrained visual modification.

    \item We evaluate ChronoPatch on four open-source VLMs, three benchmark datasets, and three commercial APIs, demonstrating latency and energy amplification on open-source models, as well as response-length and client-observed latency amplification on commercial APIs under limited perturbation and query budgets.
\end{itemize}

\section{Related Work and Preliminaries}
\label{sec:related}

\subsection{Large Vision-Language Models (VLMs)}
Modern VLMs typically combine a vision encoder with a pretrained large language model (LLM) through an alignment module, enabling autoregressive generation conditioned on visual and textual inputs. This architecture supports diverse multimodal tasks, including image captioning~\cite{li2022blip,li2023blip}, visual question answering~\cite{bai2023qwen,liu2023visual}, and multimodal reasoning~\cite{zhang2023multimodal}.

For availability analysis, autoregressive decoding is particularly relevant: the LLM generates tokens sequentially until an end-of-sequence (EOS) token is emitted or the generation budget is exhausted. Since computational cost scales approximately with the number of generated tokens~\cite{kwon2023efficient}, delayed response termination directly increases inference latency and energy consumption.

\subsection{Energy-Latency Attacks}
Energy-latency attacks target system availability by increasing inference-time resource consumption rather than degrading model accuracy. Sponge examples~\cite{shumailov2021sponge} increase neural network computation costs through inefficient activation patterns. Building on this idea, subsequent works extend resource amplification to autoregressive generation: NICGSlowDown~\cite{chen2022nicgslowdown} delays response termination in image-captioning models, while AutoDoS~\cite{zhang2025crabs} induces excessive generation in text-only LLMs.

Recent studies reveal similar vulnerabilities in VLMs. Verbose Images~\cite{gao2024inducing} uses visual perturbations to delay response termination, while RE-CITE~\cite{gao2025resource} induces repetitive generation. VLMInferSlow~\cite{wang2025vlminferslow} explores black-box energy-latency attacks using zeroth-order optimization, but assumes access to both generated text and token-level logits to guide perturbation updates. Other approaches exploit distinct attack surfaces: Overthink-Triggered Slowdown~\cite{han2026overthink} induces excessive reasoning in robotic LVLMs, Joint Pixel-Prompt Optimization~\cite{ni2026multimodal} jointly optimizes visual perturbations and prompts, and Rethinking Latency DoS~\cite{wang2026latencydos} targets serving-side KV-cache utilization and scheduler behavior.
In contrast, ChronoPatch optimizes localized visual perturbations with a fixed textual prompt, using only completed-response length as feedback, without requiring gradients, logits, or intermediate decoding signals.

\subsection{Threat Model}

\mypara{Target VLMs}
We consider autoregressive VLMs deployed as multimodal inference services. The attacker submits image-text queries to a target model and observes only final responses, with each image optimized independently under a fixed generation configuration. We evaluate open-source models for systematic analysis and commercial APIs to assess attack applicability under query-limited access.

\mypara{Adversarial Goal}
The adversary aims to compromise service availability by crafting visual perturbations that delay response termination and induce excessive generation, thereby increasing inference latency and computational resource consumption without modifying textual prompts.

\mypara{Adversary Capability}
We consider a final-response-only black-box setting in which the attacker can modify input images but observes only completed responses, using response length as the sole optimization feedback. Textual prompts and generation settings remain fixed, with no access to model internals or intermediate signals. Latency and energy are used only for evaluation. All attacks, including those on open-source models, follow this restricted access assumption. Perturbations are confined to a limited spatial region and refined to minimize their masked $\ell_2$ magnitude.

\begin{figure*}[!t]
\centering
\includegraphics[width=\textwidth]{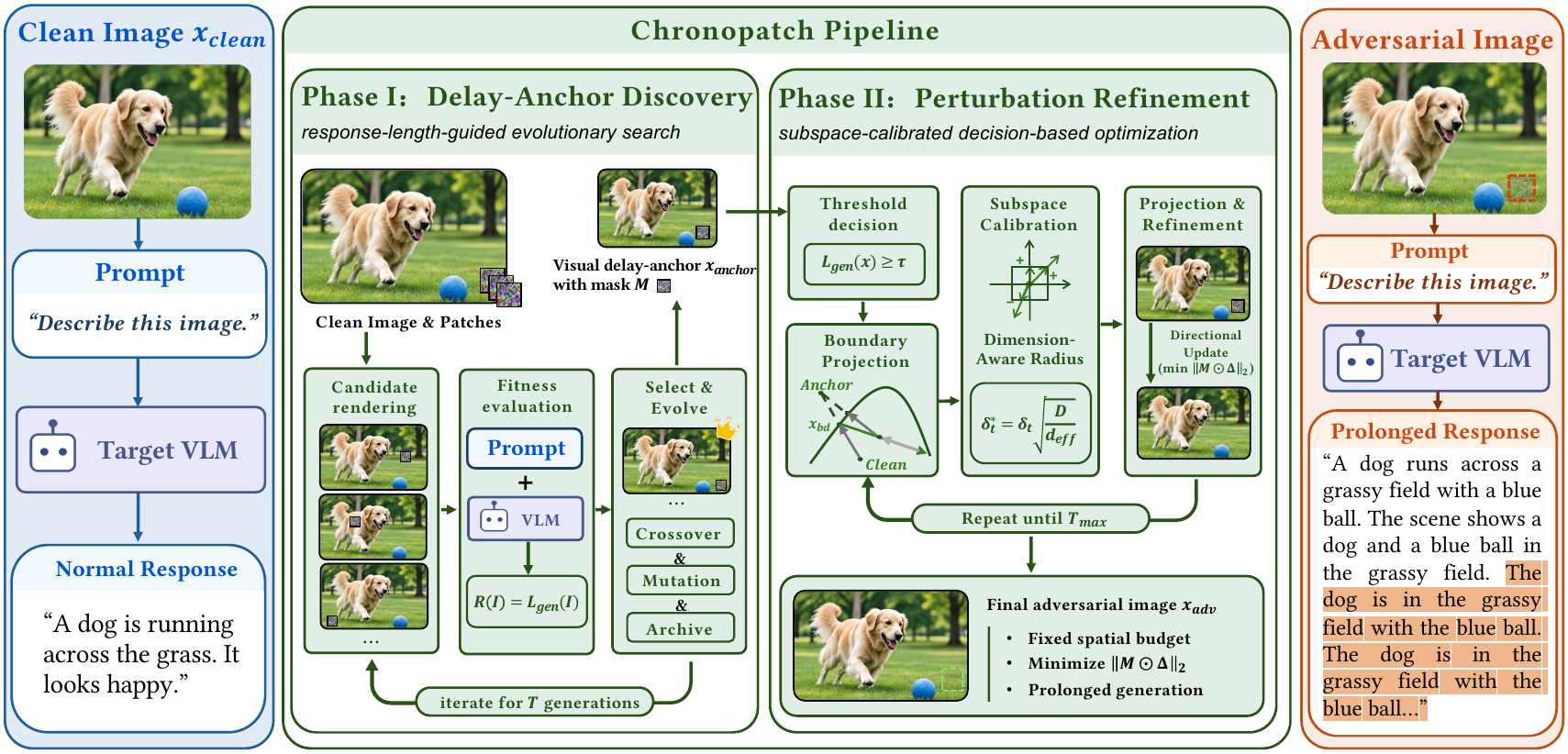}
\caption{Design overview of ChronoPatch. Phase I discovers visual delay anchors using response-length feedback. Phase II refines the anchor within a fixed spatial mask through subspace-calibrated decision-based refinement, producing a localized low-distortion adversarial image that induces longer, more repetitive responses and increases latency and energy usage.}
\label{fig:chronopatch_framework}
\end{figure*}

\section{Methodology of ChronoPatch}
\label{sec:method}

\subsection{Design Intuitions}
\label{sec:Intuitions}

ChronoPatch addresses two challenges in the final-response-only black-box setting. Since autoregressive VLMs incur additional decoding computation for each generated token, delaying response termination directly increases inference latency and energy consumption. However, without access to gradients, logits, or intermediate decoding signals, discovering inputs that induce prolonged generation and refining them under restricted visual modifications remain challenging.

\mypara{Delay-Anchor Discovery}
The first challenge is identifying visual patterns that induce prolonged generation using only final-response feedback. Although evolutionary search has been effective in discovering inputs that trigger long continuations in the text domain~\cite{guo2025prompt}, directly extending it to high-dimensional visual inputs is challenging~\cite{alzantot2019genattack}. We therefore adapt population-based exploration to visual patches, selecting and evolving candidates according to their observed response lengths. This process discovers \textit{visual delay anchors}: localized patterns that delay response termination and amplify generation cost.

\mypara{Perturbation Refinement}
The second challenge is reducing perturbation magnitude while preserving the attack effect within a restricted spatial budget. Existing decision-based methods, such as Boundary Attack~\cite{brendel2017decision} and HopSkipJumpAttack~\cite{chen2020hopskipjumpattack}, primarily explore the full image space. When only a small editable region is available, such exploration provides limited useful variation within the constrained subspace, making boundary estimation less effective. This motivates calibrating the exploration process according to the effective dimensionality of the editable region, thereby improving query efficiency.

\subsection{The ChronoPatch Framework}
As shown in \autoref{fig:chronopatch_framework}, ChronoPatch first discovers a visual delay anchor in Phase~I by jointly optimizing patch appearance and placement using response-length feedback. In Phase~II, it refines the anchor within a fixed spatial mask using binary response-length feedback, reducing perturbation magnitude while preserving prolonged generation.

\subsubsection{\textbf{Phase I: Delay-Anchor Discovery}}
Phase~I searches for a localized visual pattern that maximizes generated response length under a generation budget $G$. Given a clean image $\mathbf{x}_{\mathrm{clean}}$ and a textual prompt, we formulate anchor discovery as population-based optimization over patch appearance and placement. The complete procedure is summarized in \appalgref{alg:phase1_eogen_elitism}.

\mypara{Anchor Parameterization}
Each candidate represents a visual delay anchor composed of $N$ patches:
\begin{equation}
\mathbf{g}^{(i)}
=
\{(\mathcal{T}_{k}^{(i)}, p_{x,k}^{(i)}, p_{y,k}^{(i)})\}_{k=1}^{N},
\end{equation}
where $\mathcal{T}_{k}^{(i)} \in [0,1]^{h \times w \times c}$ denotes the $k$-th patch texture, $h$ and $w$ are the patch height and width, $c=3$ is the number of color channels, and $(p_{x,k}^{(i)},p_{y,k}^{(i)})\in[0,1]^2$ denotes its relative center location. Relative coordinates make the representation independent of input resolution.

For an input image of size $H \times W$ and spatial budget $\rho$, the patches are chosen to satisfy
\begin{equation}
Nhw \leq \rho HW.
\end{equation}
Given a candidate $\mathbf{g}^{(i)}$, the perturbed image is rendered as
\begin{equation}
\mathbf{I}^{(i)}
=
(1-\mathbf{M}^{(i)}) \odot \mathbf{x}_{\mathrm{clean}}
+
\mathbf{M}^{(i)} \odot \mathbf{P}^{(i)},
\end{equation}
where $\mathbf{M}^{(i)}$ is the binary spatial mask induced by patch locations and $\mathbf{P}^{(i)}$ is the rendered patch image. We initialize $S$ candidates with uniformly sampled textures and center-biased patch locations.

\mypara{Response-Length Reward}
Since no intermediate model feedback is available, ChronoPatch uses the generated response length as its optimization reward, 
\begin{equation}
R(\mathbf{I}) = L_{\mathrm{gen}}(\mathbf{I}),
\label{eq:RI}
\end{equation}
where $L_{\mathrm{gen}}(\mathbf{I})$ is the number of generated tokens under budget $G$. This signal is directly derived from the returned response and serves as observable feedback for delayed termination, with longer responses receiving higher rewards.

\mypara{Anchor Search}
At each iteration, ChronoPatch evaluates candidates using their generated response lengths and retains the top-$K$ candidates. The remaining candidates are generated through crossover and mutation of patch textures and locations, with annealed variances $\sigma_{\mathrm{pos}}(t)$ and $\sigma_{\mathrm{tex}}(t)$ transitioning from exploration to refinement. An archive of high-reward textures is maintained and injected with probability $P_{\mathrm{inject}}$ to preserve effective patterns. The search terminates early once a candidate reaches the generation budget $G$; otherwise, it returns the highest-reward candidate after at most $T$ iterations.

\subsubsection{\textbf{Phase II: Perturbation Refinement}}

Phase~II refines the anchor discovered in Phase~I by minimizing its perturbation magnitude while preserving prolonged generation. We optimize the perturbation values within the fixed spatial mask $\mathcal{M}$ via subspace-calibrated decision-based refinement. The procedure is summarized in \appalgref{alg:SC_hsja}.

\mypara{Constrained Refinement Objective}
Given a response-length threshold $\tau$, we define the binary decision function:
\begin{equation}
\phi(\mathbf{x}) =
\begin{cases}
+1, & L_{\mathrm{gen}}(\mathbf{x}) \geq \tau,\\
-1, & L_{\mathrm{gen}}(\mathbf{x}) < \tau,
\end{cases}
\end{equation}
where $L_{\mathrm{gen}}(\mathbf{x})$ denotes the generated response length. The threshold $\tau$ specifies the minimum response length required to preserve prolonged generation and is determined per sample using its clean response length and a model-specific ratio $\gamma_m$. This allows Phase~II to adapt to differences in benign response lengths across input samples and target models. This optimization threshold is distinct from the budget-saturation criterion used to calculate ASR in \autoref{sec:eval}.

Under the fixed spatial mask $\mathcal{M}$, Phase~II seeks the minimum perturbation satisfying the response-length constraint:
\begin{equation}
\begin{aligned}
\min_{\Delta} \quad & \|\mathcal{M}\odot \Delta\|_2 \\
\mathrm{s.t.} \quad
& \phi(\mathbf{x}_{\mathrm{clean}}+\mathcal{M}\odot\Delta)=+1, \\
& \mathbf{x}_{\mathrm{clean}}+\mathcal{M}\odot\Delta \in [0,1]^D .
\end{aligned}
\label{eq:phase2_objective}
\end{equation}
Here, $\Delta$ denotes the perturbation relative to the clean image, and $D=H\times W\times3$ is the full image dimensionality. Only perturbation components within $\mathcal{M}$ are optimized, while all other image regions remain unchanged. The objective relies on decisions derived from returned response lengths.

\mypara{Subspace-Calibrated Direction Estimation}
Standard decision-based refinement methods are typically designed for full-dimensional image optimization, whereas Phase~II operates only within the masked subspace. We define the effective dimensionality of the editable region as
\begin{equation}
d_{\mathrm{eff}} = 3\sum_{i,j}\mathcal{M}_{i,j}.
\end{equation}
Although probing directions are normalized after projection onto the editable subspace, the response-length decision boundary can be less sensitive within this restricted region. To account for the reduced effective dimensionality, ChronoPatch calibrates the probing radius as
\begin{equation}
\delta_t^{\star}
=
\delta_t \sqrt{\frac{D}{d_{\mathrm{eff}}}},
\label{eq:delta_recalibration}
\end{equation}
where $\delta_t$ is the base probing radius at iteration $t$. This dimension-based scaling provides a larger initial probing radius for smaller editable regions, while subsequent adjustments use the observed probe decisions.

At each iteration, ChronoPatch samples $B$ random directions $\{\mathbf{u}_b\}_{b=1}^{B}$ and normalizes their masked projections: $\mathbf{v}_b=(\mathcal{M}\odot\mathbf{u}_b)/\|\mathcal{M}\odot\mathbf{u}_b\|_2$. 
Probing around the current boundary point $\mathbf{x}_{\mathrm{bd}}$ yields binary labels $\phi_b\in\{-1,+1\}$, with empirical mean $\bar{\phi}=B^{-1}\sum_{b=1}^{B}\phi_b$.
When $|\bar{\phi}|$ exceeds a preset threshold, we adjust $\delta_t^{\star}$ within $[\delta_{\min},\delta_{\max}]$ until the feedback becomes balanced or a bound is reached.

Given the calibrated probe decisions, we estimate the local update direction as
\begin{equation}
\hat{\mathbf{g}}_t
=
\frac{1}{B}
\sum_{b=1}^{B}
(\phi_b-\bar{\phi})\mathbf{v}_b,
\quad
\bar{\phi}=\frac{1}{B}\sum_{b=1}^{B}\phi_b .
\label{eq:masked_gradient}
\end{equation}
Centering the probe labels reduces estimation variance, while the masked directions constrain the estimated update direction $\hat{\mathbf{g}}_t$ to the editable subspace.

\mypara{Boundary Projection and Refinement}
ChronoPatch alternates between boundary projection and masked subspace updates. Given the current adversarial image $\mathbf{x}_t$ with $\phi(\mathbf{x}_t)=+1$, it first projects the sample toward the clean image using binary search. Let $\Delta_t=\mathbf{x}_t-\mathbf{x}_{\mathrm{clean}}$. The corresponding boundary coefficient is
\begin{equation}
\alpha^{\star}
=
\inf
\left\{
\alpha\in[0,1]:
\phi\left(
\mathbf{x}_{\mathrm{clean}}+\alpha \mathcal{M}\odot\Delta_t
\right)=+1
\right\}.
\label{eq:boundary_projection}
\end{equation}

The corresponding boundary point is $\mathbf{x}_{\mathrm{bd}}=\mathbf{x}_{\mathrm{clean}}+\alpha^{\star}\mathcal{M}\odot\Delta_t$.
Starting from $\mathbf{x}_{\mathrm{bd}}$, we update the image along the estimated direction, selecting the sign that reduces its distance to the clean image. A geometric line search ensures that the updated candidate remains on the successful side of the response-length boundary:
\begin{equation}
\begin{aligned}
\mathbf{x}_{t+1}
&= \Pi_{[0,1]^D}
\left(
\mathbf{x}_{\mathrm{bd}}+\eta_t\mathbf{d}_t
\right),\\
\eta_t
&\leftarrow \eta_t/2
\quad \text{while}\quad
\phi(\mathbf{x}_{t+1})\ne +1.
\end{aligned}
\label{eq:phase2_update}
\end{equation}
where $\mathbf{d}_t \in \{\hat{\mathbf{g}}_t,-\hat{\mathbf{g}}_t\}$ is selected to reduce $\|\mathbf{x}_{t+1}-\mathbf{x}_{\mathrm{clean}}\|_2$. After $T_{\max}$ refinement iterations, we return the candidate with the smallest observed perturbation norm.

\section{Evaluation}
\label{sec:eval}

\subsection{Evaluation Setup}
\label{sec:eval_setup}

\mypara{Datasets and Target Models}
We evaluate ChronoPatch on four open-source VLMs: BLIP-Large~\cite{li2022blip}, LLaVA-v1.5-7B~\cite{liu2023visual}, Qwen3-VL-8B~\cite{bai2025qwen3}, and InternVL3-8B~\cite{zhu2025internvl3}, as well as three commercial APIs: ChatGPT-5\cite{openai2025gpt5}, Gemini-3.1-Pro-Preview~\cite{google2026gemini31pro}, and Claude-Sonnet-4-6~\cite{anthropic2026sonnet46}. We randomly sample 1,000 images from each of MSCOCO 2017~\cite{lin2014microsoft}, ImageNet~\cite{deng2009imagenet}, and Flickr30K~\cite{plummer2015flickr30k}, using semantically equivalent instructions across models.

\mypara{Baselines}
We compare ChronoPatch with Verbose Images~\cite{gao2024inducing}, a gradient-based white-box attack, and VLMInferSlow~\cite{wang2025vlminferslow}, a score-based black-box attack that leverages generated text and token-level logits. Gaussian noise~\cite{hendrycks2019benchmarking} and JPEG compression~\cite{dziugaite2016study} serve as non-adversarial controls. All methods use the same evaluation inputs and generation budgets with matched spatial perturbation constraints.

\mypara{Metrics and Success Criteria}
We evaluate attack effectiveness, perturbation distortion, query efficiency, and textual degeneration. Phase~II uses a per-sample response-length threshold $\tau$ to preserve long-generation behavior during refinement. For evaluation, we define Attack Success Rate (ASR) using a stricter budget-saturation criterion: the fraction of adversarial samples whose generated length reaches the maximum generation budget $G$, i.e., $L_{\mathrm{gen}}(\mathbf{x}_{\mathrm{adv}})\geq G$.

We measure relative increases in generated length, inference latency, and GPU energy consumption:
\begin{equation}
\begin{aligned}
\Delta L &=
\frac{L(\mathbf{x}_{\mathrm{adv}})-L(\mathbf{x}_{\mathrm{clean}})}
{L(\mathbf{x}_{\mathrm{clean}})} \times 100\%,\\
\Delta T &=
\frac{T(\mathbf{x}_{\mathrm{adv}})-T(\mathbf{x}_{\mathrm{clean}})}
{T(\mathbf{x}_{\mathrm{clean}})} \times 100\%,\\
\Delta E &=
\frac{E(\mathbf{x}_{\mathrm{adv}})-E(\mathbf{x}_{\mathrm{clean}})}
{E(\mathbf{x}_{\mathrm{clean}})} \times 100\%,
\end{aligned}
\end{equation}
where $L$, $T$, and $E$ denote generated token length, inference latency, and GPU energy consumption, respectively. Perturbation distortion is measured using the masked $\ell_2$ norm and LPIPS~\cite{zhang2018unreasonable}. We report stage-wise query counts $Q_{\mathrm{I}}$ and $Q_{\mathrm{II}}$, with $Q_{\mathrm{total}}=Q_{\mathrm{I}}+Q_{\mathrm{II}}$. Each image-text model evaluation counts as one query, regardless of batching.

We quantify textual degeneration using the $n$-gram repetition ratio~\cite{welleck2019neural,su2022contrastive}:
\begin{equation}
\mathrm{rep}\text{-}n
=1-\frac{|\mathrm{unique}\ n\text{-grams}|}
{|\mathrm{total}\ n\text{-grams}|}.
\end{equation}
We report \rep{2}, \rep{3}, and \rep{4}, using \rep{4} as the primary measure of longer-span repetition.

\begin{table*}[!t]
\caption{Main experimental results across four target VLMs on MSCOCO and ImageNet. $\Delta$ L, $\Delta$ T, and $\Delta$ E represent the percentage increase compared to benign clean-image inference. Best and second-best results are highlighted in \textbf{bold} and \underline{underlined}, respectively.}
\label{tab:main_results}
\centering
\resizebox{\textwidth}{!}{
\begin{tabular}{ll cccc cccc}
\toprule
\multirow{2}{*}{\textbf{Target Models}} & \multirow{2}{*}{\textbf{Attack Method}} & \multicolumn{4}{c}{\textbf{MSCOCO}} & \multicolumn{4}{c}{\textbf{ImageNet}} \\
\cmidrule(lr){3-6} \cmidrule(lr){7-10}
& & \textbf{Avg. Len} & \textbf{$\Delta$ L (\%)} & \textbf{$\Delta$ T (\%)} & \textbf{$\Delta$ E (\%)} & \textbf{Avg. Len} & \textbf{$\Delta$ L (\%)} & \textbf{$\Delta$ T (\%)} & \textbf{$\Delta$ E (\%)} \\
\midrule

\multirow{5}{*}{BLIP-Large}
& JPEG & 22.75 & 0.31\% & -12.60\% & -8.41\% & 21.90 & 2.29\% & 1.47\% & 3.23\% \\
& Gaussian & 22.32 & -1.59\% & -13.31\% & -8.91\% & 20.73 & -3.18\% & -3.35\% & -1.05\% \\
& Verbose Images & 23.41 & 4.26\% & 4.90\% & 5.52\% & 23.67 & 10.56\% & 11.43\% & 9.60\% \\
& VLMInferSlow & \underline{25.85} & \underline{13.99\%} & \underline{11.49\%} & \underline{10.65\%} & \underline{26.15} & \underline{22.14\%} & \underline{18.77\%} & \underline{21.51\%} \\
& ChronoPatch (Ours) & \textbf{29.92} & \textbf{47.92\%} & \textbf{47.86\%} & \textbf{41.44\%} & \textbf{30.27} & \textbf{41.38\%} & \textbf{40.91\%} & \textbf{34.36\%} \\

\midrule

\multirow{5}{*}{LLaVA-v1.5-7B}
& JPEG & 127.16 & 0.30\% & 1.24\% & 2.13\% & 119.89 & 11.79\% & 13.60\% & 12.38\% \\
& Gaussian & 123.81 & -2.93\% & -3.85\% & -2.59\% & 118.42 & 10.41\% & 9.27\% & 9.44\% \\
& Verbose Images & \textbf{201.67} & \textbf{58.11\%} & \textbf{54.19\%} & \textbf{49.31\%} & \underline{192.26} & \underline{79.26\%} & \underline{80.31\%} & \underline{73.08\%} \\
& VLMInferSlow & 158.75 & 25.34\% & 25.13\% & 24.87\% & 163.52 & 52.47\% & 55.53\% & 56.53\% \\
& ChronoPatch (Ours) & \underline{185.22} & \underline{45.21\%} & \underline{47.08\%} & \underline{43.89\%} & \textbf{197.38} & \textbf{84.04\%} & \textbf{85.20\%} & \textbf{81.55\%} \\

\midrule

\multirow{5}{*}{Qwen3-VL-8B}
& JPEG & 397.02 & -1.60\% & -1.77\% & -1.27\% & 364.48 & -6.10\% & -5.75\% & -5.14\% \\
& Gaussian & 377.16 & -6.52\% & -6.64\% & -4.78\% & 372.51 & -4.03\% & -6.11\% & -3.52\% \\
& Verbose Images & \underline{726.86} & \underline{84.02\%} & \underline{83.54\%} & \underline{60.15\%} & \underline{695.04} & \underline{79.07\%} & \underline{81.92\%} & \underline{68.98\%} \\
& VLMInferSlow & 487.77 & 20.89\% & 19.22\% & 23.84\% & 517.93 & 33.45\% & 30.29\% & 31.81\% \\
& ChronoPatch (Ours) & \textbf{938.13} & \textbf{132.53\%} & \textbf{142.42\%} & \textbf{129.56\%} & \textbf{962.58} & \textbf{148.91\%} & \textbf{151.20\%} & \textbf{138.96\%} \\

\midrule

\multirow{5}{*}{InternVL3-8B}
& JPEG & 252.70 & 3.71\% & 0.33\% & 0.28\% & 230.11 & -0.07\% & 0.73\% & 1.62\% \\
& Gaussian & 225.01 & -7.66\% & -0.15\% & -4.39\% & 216.58 & -8.29\% & -6.35\% & -5.40\% \\
& Verbose Images & \underline{388.73} & \underline{59.54\%} & \underline{67.86\%} & \underline{57.68\%} & \underline{397.61} & \underline{68.37\%} & \underline{71.03\%} & \underline{60.38\%} \\
& VLMInferSlow & 321.27 & 31.85\% & 28.01\% & 33.81\% & 335.29 & 41.98\% & 37.15\% & 45.58\% \\
& ChronoPatch (Ours) & \textbf{1022.36} & \textbf{319.56\%} & \textbf{293.34\%} & \textbf{249.70\%} & \textbf{1021.96} & \textbf{332.76\%} & \textbf{279.61\%} & \textbf{237.67\%} \\

\bottomrule
\end{tabular}
}
\end{table*}

\mypara{Implementation and Measurement}
Experiments use PyTorch on NVIDIA RTX PRO 6000 GPUs (96 GB). Open-source models employ greedy decoding without repetition controls, with generation budgets of $G=512$ for BLIP-Large and $G=1024$ for other models, under a 5\% spatial budget. Phase~I initializes $S=100$ candidates and performs up to $T=500$ search iterations; Phase~II performs up to $T_2=50$ refinement iterations with $B=128$ probes per iteration. We report mean latency over three batch-size-one runs on a dedicated GPU, excluding preprocessing and synchronizing GPU execution; GPU energy is measured by NVML. For commercial APIs, randomized clean/adversarial request pairs use identical prompts and user-controllable settings. We report provider completion-token counts and client-observed end-to-end latency. Additional hyperparameters are provided in \autoref{app:implementation}.

\subsection{Main Results}

\mypara{Effectiveness against Baselines}
As shown in \autoref{tab:main_results}, ChronoPatch consistently increases generated length, inference latency, and GPU energy consumption across all evaluated models. On MSCOCO, latency increases by 47.86\%--293.34\% and GPU energy by 41.44\%--249.70\%. On ImageNet, the corresponding increases are 40.91\%--279.61\% and 34.36\%--237.67\%, respectively. InternVL3 exhibits the strongest amplification, with average response lengths approaching the 1,024-token generation limit on both datasets.

ChronoPatch substantially outperforms non-adversarial perturbations and achieves greater latency amplification than VLMInferSlow across all evaluated settings, despite relying only on final-response feedback. It also exceeds the white-box Verbose Images baseline in seven of eight model--dataset settings. The concurrent increases are consistent with the additional cost of prolonged autoregressive decoding. These results demonstrate substantial resource amplification under a final-response-only black-box setting.

\mypara{Repetitive Generation}
ChronoPatch also increases output repetition, as measured by the $n$-gram repetition ratios in \autoref{tab:efficiency_degeneration}. While BLIP and LLaVA exhibit relatively modest increases in \rep{4}, Qwen3-VL shows an increase from 0.39\% to 29.54\%, and InternVL3 from 8.07\% to 65.81\%. The corresponding increases in \rep{2} and \rep{3} further indicate that this repetition occurs across multiple n-gram orders. These results show that increased generation length is accompanied by substantial repetitive degeneration, particularly on Qwen3-VL and InternVL3.

\begin{table}[!htbp]
\centering
\caption{Generated-length increase and $n$-gram repetition ratios on MSCOCO. Higher \rep{4}  indicates more severe repetition.}
\label{tab:efficiency_degeneration}
\resizebox{\columnwidth}{!}{
\begin{tabular}{l c c c c c}
\toprule
\multirow{2}{*}{\textbf{Models}} 
& \multirow{2}{*}{\textbf{Setup}} 
& \multirow{2}{*}{\textbf{$\Delta$ L (\%)$\uparrow$}} 
& \multicolumn{3}{c}{\textbf{N-gram Repetition Ratio (\%)}} \\

\cmidrule(l){4-6}

& & 
& \textbf{$rep\text{-}2$ $\uparrow$} 
& \textbf{$rep\text{-}3$ $\uparrow$} 
& \textbf{$rep\text{-}4$ $\uparrow$} \\

\midrule

\multirow{2}{*}{BLIP} 
& Clean & \multirow{2}{*}{+47.92\%} & 0.21 & 0.08 & 0.01 \\
& Adv & & 4.72 & 3.37 & 2.58 \\

\midrule

\multirow{2}{*}{LLaVA} 
& Clean & \multirow{2}{*}{+45.21\%} & 6.49 & 1.21 & 0.42 \\
& Adv & & 14.67 & 5.84 & 2.93 \\

\midrule

\multirow{2}{*}{Qwen3-VL} 
& Clean & \multirow{2}{*}{+110.68\%} & 8.69 & 1.50 & 0.39 \\
& Adv & & 37.09 & 30.86 & 29.54 \\

\midrule

\multirow{2}{*}{InternVL3} 
& Clean & \multirow{2}{*}{+319.56\%} & 13.66 & 10.11 & 8.07 \\
& Adv & & 71.86 & 67.84 & 65.81 \\

\bottomrule
\end{tabular}
}
\end{table}

\mypara{Query Cost and Refinement}
We analyze the trade-off between query cost, attack effectiveness, and perturbation magnitude under the 5\% spatial budget. In Phase~I, query caps of 20K, 30K, and 40K retain 91.85\%, 92.90\%, and 91.16\% of the final response-length gains on LLaVA, Qwen3-VL, and BLIP, respectively, indicating diminishing returns from further anchor search. Phase~II progressively reduces masked $\ell_2$ distortion with additional queries, exhibiting model-dependent convergence. InternVL3 converges fastest, reducing its $\ell_2$ norm from 4.8 at approximately 1K queries to 1.56 near 8K, whereas Qwen3-VL decreases from 20.0 to 4.5 over a comparable range. Mean LPIPS values range from 0.0105 to 0.0188, with 90th-percentile values below 0.033. Visual examples are shown in \autoref{fig:visual_comparison}, with detailed convergence and distortion results provided in the Supplementary Material.

\begin{figure}[t]
\centering
\includegraphics[width=\columnwidth]{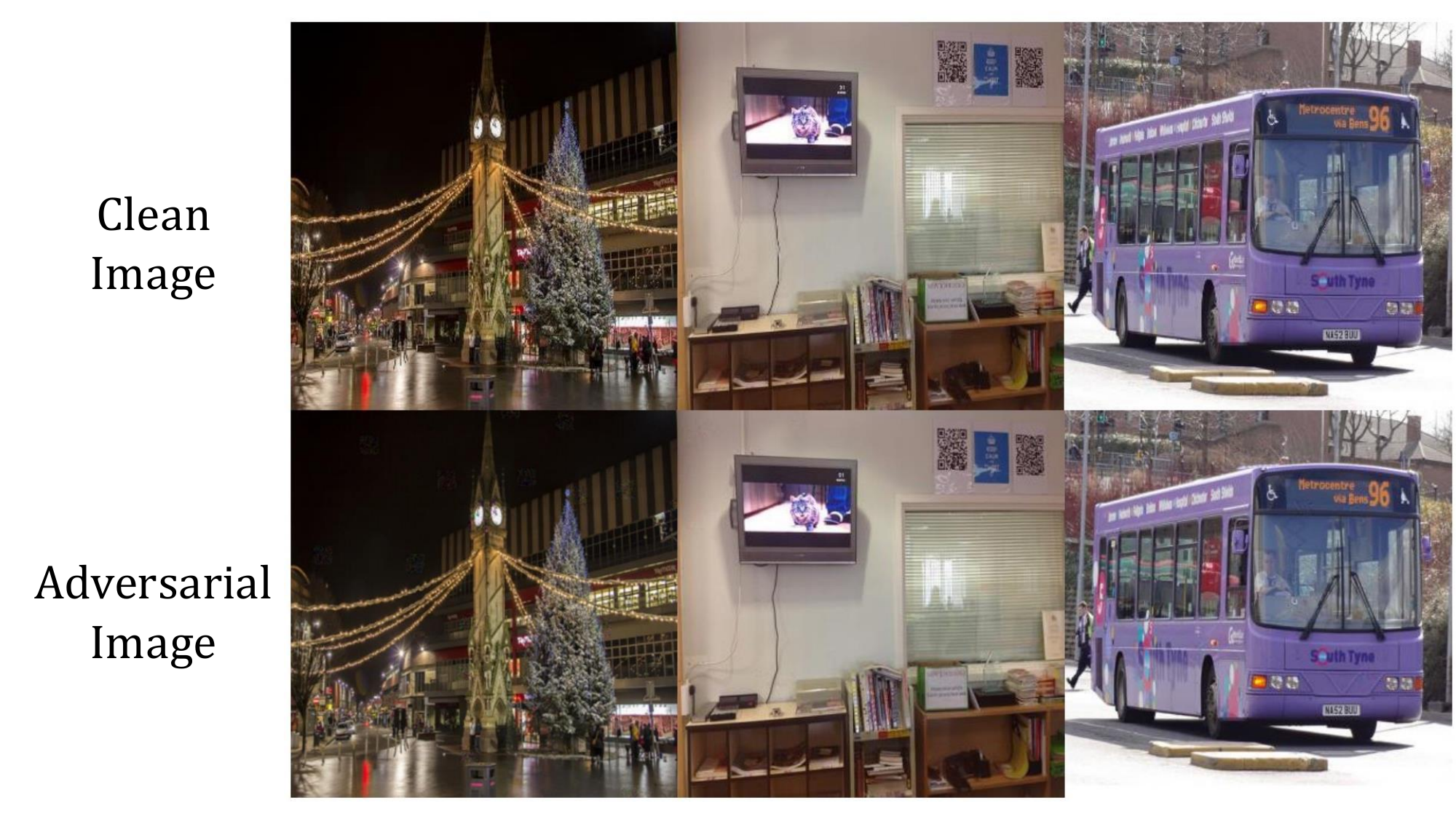}
\caption{Visual comparison of clean images (top) and their corresponding adversarial images (bottom).}
\label{fig:visual_comparison}
\end{figure}

\subsection{Ablation Study}

\mypara{Ablation on Delay-Anchor Discovery}
We evaluate three Phase~I components on MSCOCO 2017 using Qwen3-VL: texture archive injection, annealed mutation, and center-biased initialization. Here, ASR denotes the fraction of samples for which Phase~I discovers at least one candidate reaching the generation budget $G$. As shown in \autoref{tab:ablation_study_I}, the complete Phase~I configuration achieves 95.8\% ASR with an average of 961.22 queries. Removing any component reduces ASR to 75.13\%--78.49\% and increases the average query count to 1,139.36--1,336.29. These results highlight the contribution of each component to reliable and query-efficient delay-anchor discovery. We further evaluate two generic black-box search baselines, a vanilla genetic algorithm and random search. They achieve ASRs of 66.7\% and 63.3\%, with average response lengths of 738.77 and 729.63 tokens, respectively, demonstrating that response-length feedback can also guide generic search methods toward effective delay anchors.

\begin{table}[!htbp]
\centering
\caption{Ablation study of the Phase~I delay-anchor discovery components on MSCOCO 2017 with Qwen3-VL.}
\label{tab:ablation_study_I}
\resizebox{\columnwidth}{!}{
\begin{tabular}{l c c c}
\toprule
\textbf{Method Variants} & \textbf{ASR (\%)} $\uparrow$ & \textbf{Avg. Length} $\uparrow$ & \textbf{Avg. Queries} $\downarrow$ \\
\midrule
Full Method (Ours)       & \textbf{95.8} & \textbf{938.13} & \textbf{961.22} \\
\midrule
w/o Texture Archive           & 75.13 & 743.33 & 1298.33\\
w/o Annealed Mutation   & 78.49 & 759.29 & 1139.36\\
w/o Center-biased Init.        & 75.56 & 739.46 & 1336.29\\
\bottomrule
\end{tabular}
}
\end{table}

\mypara{Ablation on Perturbation Refinement}
We compare Phase~II with a refinement baseline based on HopSkipJumpAttack (HSJA), a standard decision-based black-box optimization method. Both methods use the same initial delay anchors, spatial masks, and query budgets, while the HSJA baseline does not apply dimensionality-aware probe calibration. As shown in \autoref{tab:trajectory_comparison}, ChronoPatch consistently achieves lower perturbation norms across all four VLMs. At 1,000 queries, ChronoPatch achieves masked $\ell_2$ norms of 10.36 and 4.14 on Qwen3-VL and InternVL3, respectively, compared with 25.07 and 22.15 for the HSJA baseline. These results support the effectiveness of subspace-calibrated refinement under spatial constraints.

\begin{table}[!htbp]
\centering
\caption{Comparison of the $\ell_2$ perturbation norm at specific query intervals between Standard HSJA and our perturbation refinement method under spatial constraints.}
\label{tab:trajectory_comparison}
\resizebox{\columnwidth}{!}{
\begin{tabular}{l l c c c }
\toprule
\textbf{Method} & \textbf{Models} & \textbf{Query 100} & \textbf{Query 500} & \textbf{Query 1000}  \\
\midrule
\multirow{4}{*}{Standard HSJA} 
 & BLIP & 27.04 & 23.13 & 21.15 \\
 & LLaVA & 28.19 & 24.40 & 20.61 \\
 & Qwen3-VL & 27.45 & 25.13 & 25.07 \\
 & InternVL3 & 26.73 & 22.98 & 22.15 \\
\midrule
\multirow{4}{*}{\shortstack[l]{Perturbation\\Refinement (Ours)}}
 & BLIP & 21.69 & 13.22 & 8.51  \\
 & LLaVA & 20.11 & 14.92 & 9.67  \\
 & Qwen3-VL & 22.50 & 16.07 & 10.36  \\
 & InternVL3 & 18.71 & 10.11 & 4.14  \\
\bottomrule
\end{tabular}
}
\end{table}

\begin{figure}[!htbp]
\centering
\includegraphics[width=\columnwidth]{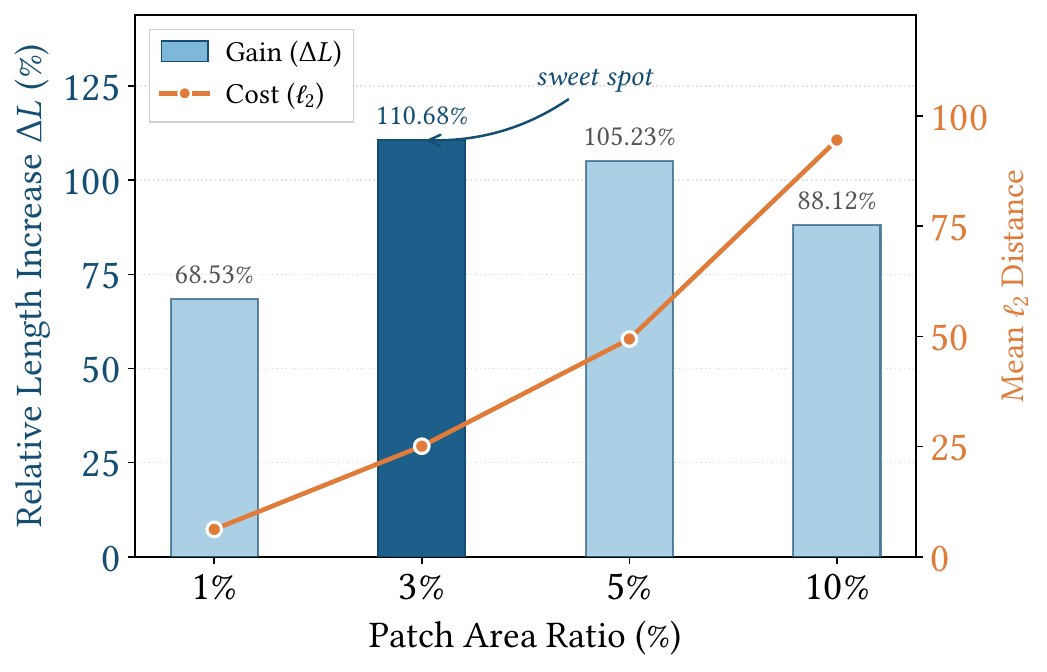}
\caption{Effect of spatial budget on Qwen3-VL. Bars show relative response-length increase ($\Delta L$), while the line shows the masked $\ell_2$ perturbation norm.}
\label{fig:patch_size_ablation}
\end{figure}

\mypara{Sensitivity to Spatial Budget}
We investigate the effect of spatial budget by varying the editable area from 1\% to 10\% on Qwen3-VL. As shown in \autoref{fig:patch_size_ablation}, ChronoPatch increases generation length by 68.53\% with only a 1\% budget, while a 3\% budget yields 110.68\% with a masked $\ell_2$ norm of 25.10. Increasing the budget to 10\% reduces $\Delta L$ to 88.12\% while raising the norm to 94.57. These results reveal a non-monotonic relationship between spatial budget and attack effectiveness, with the 3\%--5\% range offering a favorable balance between generation amplification and perturbation magnitude.

\subsection{Additional Analysis}

\mypara{Effect of Model Scale}
We evaluate ChronoPatch across four Qwen3-VL variants ranging from 2B to 32B parameters. Here, $Q_{\mathrm{break}}$ denotes the average number of queries required for an adversarial sample to reach the maximum generation budget $G$, while ASR@500 measures the success rate under a fixed 500-query budget. As shown in \autoref{tab:model_scale}, all variants achieve final $\Delta L$ values exceeding 110\%. However, query costs increase substantially with model size: the average $Q_{\mathrm{break}}$ rises from 311.89 for the 2B variant to 1092.71 for the 32B variant, while ASR@500 decreases from 92.32\% to 35.60\%. These results indicate that larger variants require more queries to discover effective delay anchors, although substantial length amplification remains achievable. 

\begin{table}[!htbp]
\centering
\caption{Effect of model scale across Qwen3-VL variants. $Q_{\mathrm{break}}$ denotes the average queries required to reach the generation budget $G$, while ASR@500 denotes the success rate at 500 queries.}
\label{tab:model_scale}
\resizebox{\columnwidth}{!}{
\begin{tabular}{l c c c}
\toprule
\textbf{Model} & \textbf{Final $\Delta L$ (\%)} & \textbf{Avg. $Q_{\mathrm{break}}$ ($\downarrow$)} & \textbf{ASR@500} \\
\midrule
Qwen3-VL-2B  & +182.79\% & \textbf{311.89}  & \textbf{92.32\%} \\
Qwen3-VL-4B  & +171.55\% & 461.22         & 82.41\% \\
Qwen3-VL-8B  & +132.53\% & 793.89         & 68.59\% \\
Qwen3-VL-32B & +110.12\% & 1092.71        & 35.60\% \\
\bottomrule
\end{tabular}
}
\end{table}

\mypara{Robustness and Defense Evaluation}
We evaluate two generation-time defenses on 100 held-out clean/adversarial pairs per model (Qwen3-VL and InternVL3), using fixed Phase~I attack images. Repetition penalty ($r=1.2$) reduces mean adversarial length by 45.40\% and 66.27\%, respectively, but also alters 94 and 51 clean responses, with mean length changes of $-3.91\%$ and $+7.00\%$. Repetition-aware early stopping (128-token window, 30\% $\mathrm{rep}\text{-}4$ threshold) reduces adversarial length by 26.77\% and 36.02\%, truncating 29 and 26 attack responses. We also evaluate four input transformations (JPEG compression~\cite{dziugaite2016study}, Gaussian blur~\cite{hendrycks2019benchmarking}, random resizing and padding~\cite{xie2017mitigating}, and feature quantization~\cite{xu2017feature}) on Qwen3-VL. These transformations reduce $\Delta L$ from 110.68\% to 45.48\%--67.88\% and $\Delta T$ from 142.42\% to 49.65\%--74.57\%, yet substantial amplification persists. Overall, both generation-time defenses mitigate but do not fully eliminate attack-induced over-generation, highlighting trade-offs between attack suppression and interference with benign generation.

\subsection{Evaluation on Commercial APIs}
\label{sec:commercial_api_eval}

We further evaluate ChronoPatch on three closed-source commercial VLM APIs with a total query budget of at most $Q_{\mathrm{total}}\leq 500$. 
As shown in \autoref{tab:commercial_api}, ChronoPatch increases generated length by 37.6\%--81.5\% and end-to-end latency by 1.82$\times$ to 2.29$\times$ across the three commercial services. These results demonstrate that over-generation remains observable in deployed black-box VLM services.

\begin{table}[h]
\centering
\caption{Evaluation on closed-source commercial APIs under $Q_{\mathrm{total}}\leq500$ queries per API. $\Delta L$ and $\Delta T$ denote relative increases in response length and client-observed end-to-end latency.}
\label{tab:commercial_api}
\resizebox{\columnwidth}{!}{
\begin{tabular}{l c c c}
\toprule
\textbf{Commercial API} & \textbf{Query Budget} & \textbf{$\Delta L$ (\%)} & \textbf{$\Delta T$ ($\times$)}  \\
\midrule
ChatGPT-5 & $\le 500$ & 81.5\% & 2.25$\times$  \\
Gemini-3.1-Pro-Preview & $\le 500$ & 42.4\% & 1.82$\times$  \\
Claude-Sonnet-4-6 & $\le 500$ & 37.6\% & 2.29$\times$  \\
\bottomrule
\end{tabular}
}
\end{table}

\section{Discussion}
\label{sec:discussion}

\mypara{Attribution Analysis}
Grad-CAM~\cite{selvaraju2017grad} visualizations on BLIP-Large (see Supplementary Material) reveal shifts in highlighted regions toward adversarial patches, providing qualitative evidence of altered visual attribution rather than a causal explanation of prolonged generation.

\mypara{Limitations}
Our main attack evaluation uses controlled decoding configurations, including greedy decoding and fixed generation budgets. ChronoPatch also incurs substantial per-image optimization costs, particularly during extended Phase~I search and Phase~II refinement. However, optimized adversarial images may be reused across repeated requests under matched target conditions, potentially amortizing the one-time optimization cost. For commercial APIs, we measure client-observed end-to-end latency, while GPU energy measurements are limited to open-source models. Although the perturbations are spatially localized and exhibit low image-level LPIPS, their local appearance may remain perceptible; improving their semantic blending is an important direction for future work.

\section{Conclusion}
\label{sec:conclusion}

This paper presents ChronoPatch, a final-response-only black-box energy-latency attack against vision-language models. Using response length as its sole feedback signal, ChronoPatch combines delay-anchor discovery with subspace-calibrated perturbation refinement to identify localized visual perturbations that induce prolonged autoregressive generation. Experiments across four open-source VLMs show consistent increases in response length, inference latency, and GPU energy consumption, accompanied by repetitive degeneration in generated responses. Evaluations on three commercial APIs further show that the over-generation effect remains observable under a total query budget of at most 500 queries per API. These findings show that restricting access to intermediate model feedback alone does not eliminate generation-based availability risks, underscoring the need for deployment-level safeguards that monitor abnormal generation and combine input preprocessing with generation-aware decoding controls.

\bibliographystyle{plainurl}
\bibliography{cite}

\clearpage
\appendix
\raggedbottom
\newenvironment{chronoappendixfigure}{%
  \par\addvspace{\medskipamount}\noindent
  \begin{minipage}{\columnwidth}
  \captionsetup{type=figure,position=bottom,skip=4pt}\centering
}{\end{minipage}\par\addvspace{\medskipamount}}
\newenvironment{chronoappendixtable}{%
  \par\addvspace{\medskipamount}\noindent
  \begin{minipage}{\columnwidth}
  \captionsetup{type=table,position=top,skip=4pt}\centering
}{\end{minipage}\par\addvspace{\medskipamount}}

\section{Implementation Details}
\label{app:implementation}

\mypara{Phase I Hyperparameters}
\autoref{tab:phase1_hyperparams} summarizes the key hyperparameters used in the delay-anchor discovery stage. The candidate set size is set to $S = 100$ with a maximum of $T = 500$ anchor-search iterations. Each patch texture $\mathcal{T}$ is initialized by uniform sampling from $[0,1]^{h \times w \times c}$, and patch locations are sampled from a Gaussian distribution centered at the image center with standard deviation 0.2 (in relative coordinates), reflecting the prior that semantically salient regions tend to concentrate near the image center~\cite{liu2010learning}. Mutation variances decay linearly from $\sigma = 0.1$ to $0.02$ over the course of optimization, and the archived texture injection probability is fixed at $P_{\text{inject}} = 0.15$.
In experiments involving closed-source VLM APIs, we use conservative settings for candidate set size, maximum iterations, and retained candidate count to respect query and cost constraints.

\begin{chronoappendixtable}
\centering
\caption{Hyperparameter settings for Phase I (Delay-Anchor Discovery).}
\label{tab:phase1_hyperparams}
\resizebox{\columnwidth}{!}{
\begin{tabular}{lcc}
\toprule
\textbf{Hyperparameter} & \textbf{Symbol} & \textbf{Value} \\
\midrule
Candidate set size      & $S$             & 100            \\
Max anchor-search iterations & $T$        & 500            \\
Retained candidate count & $K$            & 32             \\
Initial mutation variance & $\sigma_{\max}$ & 0.1          \\
Final mutation variance & $\sigma_{\min}$ & 0.02           \\
Archived texture injection probability & $P_{\text{inject}}$ & 0.15   \\
Max generation budget   & $G$             & 512 / 1024     \\
\bottomrule
\end{tabular}
}
\end{chronoappendixtable}

\mypara{Phase II Hyperparameters}
\autoref{tab:phase2_ratios} summarizes the model-specific decision ratios used to set response-length thresholds during perturbation refinement. Because different VLMs exhibit substantially different benign response lengths, we use a model-specific ratio $\gamma_m$ to derive the token threshold $\tau_m$ from clean-output length statistics. The resulting threshold is chosen above the typical benign generation range but below the response length induced by the Phase-I delay anchor. This ensures that Phase II preserves a meaningful over-generation effect while allowing sufficient slack for perturbation compression. If the derived threshold exceeds the maximum generation budget $G$, we cap it at $G$, i.e., $\tau_m \leftarrow \min(\tau_m, G)$.

\begin{chronoappendixtable}
\centering
\caption{Model-specific decision ratios used to set Phase II response-length thresholds.}
\label{tab:phase2_ratios}
\begin{tabular}{lc}
\toprule
\textbf{Target Model} & \textbf{Decision Ratio $\gamma_m$} \\
\midrule
BLIP-Large       & 1.30 \\
LLaVA-v1.5-7B    & 1.30 \\
Qwen3-VL-8B      & 2.00 \\
InternVL3-8B     & 2.00 \\
\bottomrule
\end{tabular}
\end{chronoappendixtable}

The remaining Phase-II hyperparameters are shared across target models and are reported in \autoref{tab:phase2_hyperparams}. The maximum number of iterations is set to $T_{\max}=50$, with a direction-estimation batch size of $B=128$. The trust-region dynamic scaling is triggered when the mean probe feedback satisfies $|\mathbb{E}[\phi_b]|\geq0.95$, scaling the calibrated radius $\delta_t^*$ by a factor of 3.0 or 0.5 accordingly.

\begin{chronoappendixtable}
\centering
\caption{Shared hyperparameter settings for Phase II (Perturbation Refinement).}
\label{tab:phase2_hyperparams}
\begin{tabular}{lcc}
\toprule
\textbf{Hyperparameter} & \textbf{Symbol} & \textbf{Value} \\
\midrule
Max iterations                  & $T_{\max}$       & 50       \\
Direction estimation batch size & $B$              & 128      \\
Trust-region scale-up factor   & ---              & 3.0      \\
Trust-region scale-down factor & ---              & 0.5      \\
Trigger threshold              & ---              & 0.95     \\
$\delta$ lower bound           & $\delta_{\min}$  & $10^{-4}$\\
$\delta$ upper bound           & $\delta_{\max}$  & $10^{0}$ \\
\bottomrule
\end{tabular}
\end{chronoappendixtable}

\mypara{Compute Infrastructure}
All experiments are conducted on four NVIDIA RTX PRO 6000 GPUs (96GB VRAM) using PyTorch 2.x. Commercial API experiments are conducted with a strict query budget of $Q_{\max} \leq 500$ to minimize billing costs and respect rate limits.

\section{Additional Experimental Results}
\label{app:additional_results}

\mypara{Generated-Length Distributions}
\autoref{fig:distribution} compares response-length distributions for clean and adversarial inputs across BLIP, LLaVA, and Qwen3-VL. The adversarial distributions shift toward longer generations, complementing the aggregate results in \autoref{tab:main_results}.

\mypara{Query Efficiency and Optimization Behavior}
\autoref{fig:progress} illustrates the evolution of response length during Phase~I, showing the progression of delay-anchor discovery across target models. \autoref{fig:query} reports the masked $\ell_2$ norm as a function of Phase~II queries, illustrating the progressive reduction of perturbation magnitude. Together, these results characterize the query costs and optimization behavior of the two stages.

\mypara{Perturbation Magnitude and Perceptual Distortion}
\autoref{tab:lpips} reports masked $\ell_2$ norms and LPIPS scores after 1K Phase~II queries. Mean LPIPS ranges from 0.0105 to 0.0188 across the four models, with all 90th-percentile values below 0.033. These results indicate low image-level perceptual distortion under the 5\% spatial budget.

\begin{chronoappendixtable}
\centering
\caption{Perturbation magnitude and perceptual distortion of ChronoPatch after 1,000 refinement queries under the default 5\% spatial budget on MSCOCO. $\ell_2$ is computed over the masked perturbation, and LPIPS is computed between clean and adversarial images. Lower values indicate smaller perturbations or higher perceptual similarity.}
\label{tab:lpips}
\resizebox{\columnwidth}{!}{
\begin{tabular}{l c c c}
\toprule
\textbf{Models} & \textbf{$\ell_2$ @ 1K Queries}   & \textbf{Mean LPIPS} $\downarrow$ & \textbf{P90 LPIPS} $\downarrow$ \\
\midrule
BLIP   & 8.51 & 0.0137 & 0.0298 \\
LLaVA   & 9.67 & 0.0141 & 0.0307 \\
Qwen3-VL     & 10.36 & 0.0188 & 0.0324 \\
InternVL3    & 4.14 & 0.0105 & 0.0211 \\
\bottomrule
\end{tabular}
}
\end{chronoappendixtable}

\twocolumn[{%
\begin{minipage}{\textwidth}
\begingroup\captionsetup{type=figure}
\centering

\includegraphics[width=0.32\textwidth]{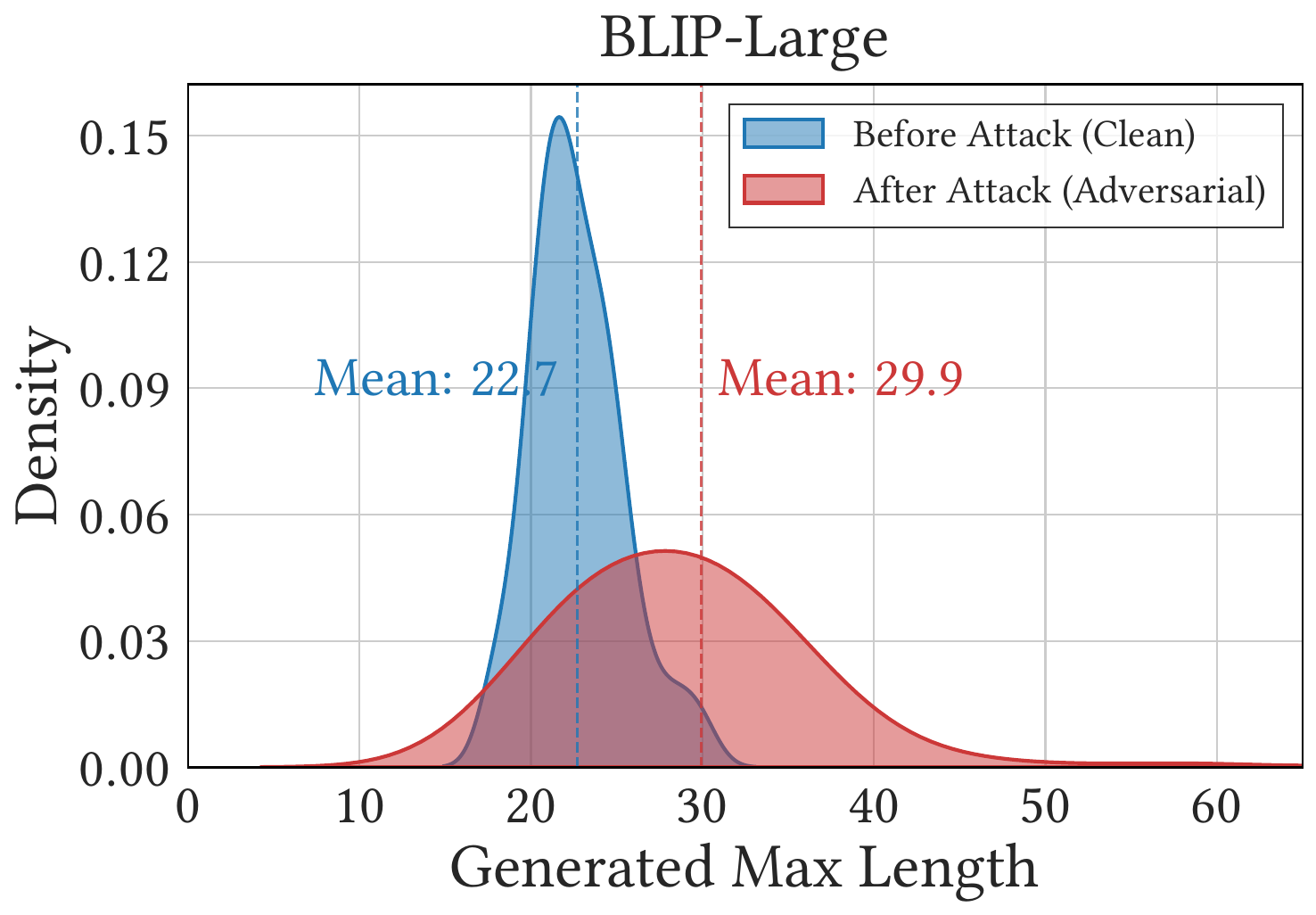}
\hfill
\includegraphics[width=0.32\textwidth]{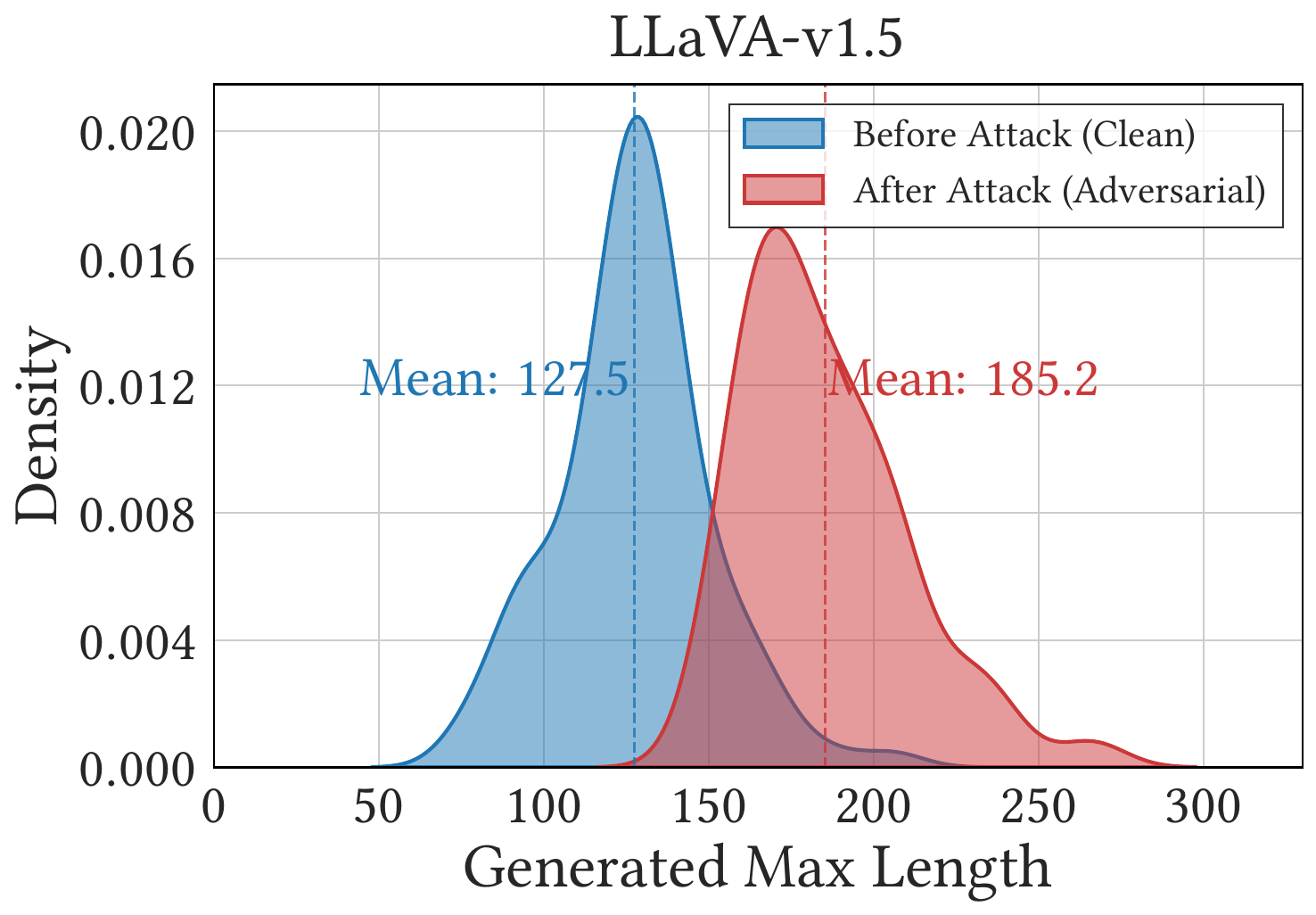}
\hfill
\includegraphics[width=0.32\textwidth]{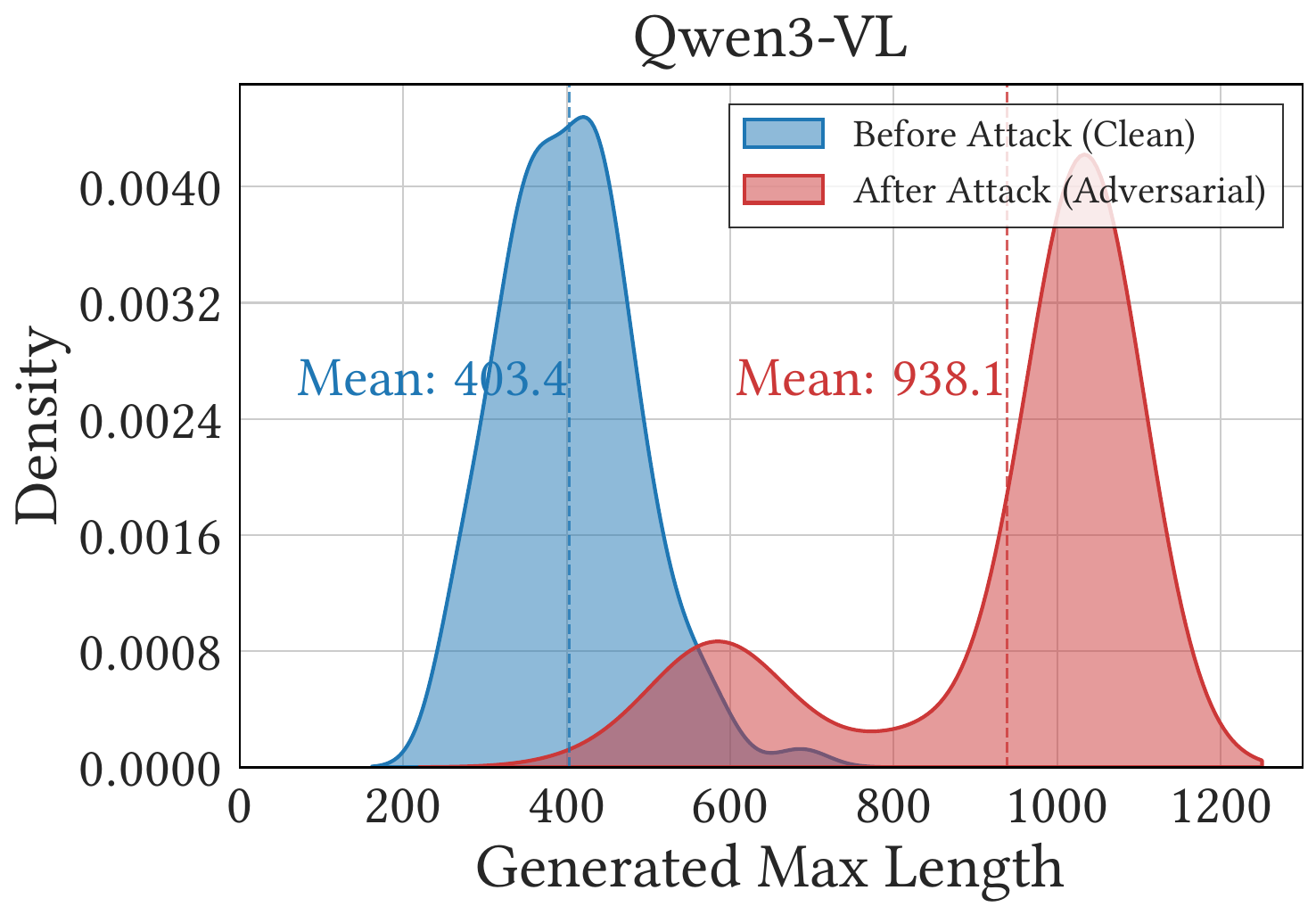}
\caption{Distributions of generated sequence lengths for clean and adversarial images on representative target models. The adversarial distributions shift toward longer generations, indicating that ChronoPatch consistently increases response length across samples.}
\label{fig:distribution}
\endgroup
\medskip
\begingroup\captionsetup{type=figure}
\centering
\includegraphics[width=0.32\textwidth]{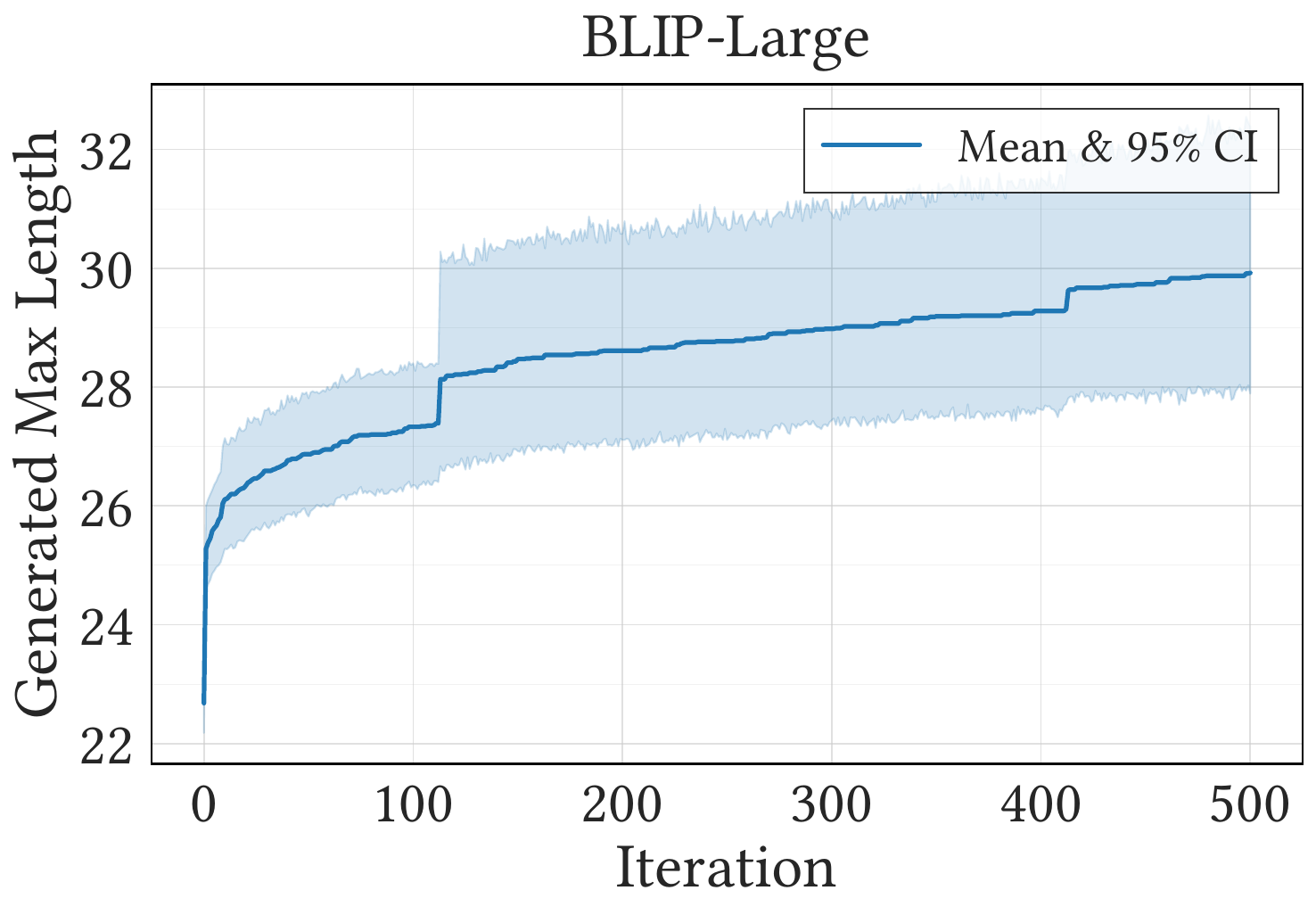}
\hfill
\includegraphics[width=0.32\textwidth]{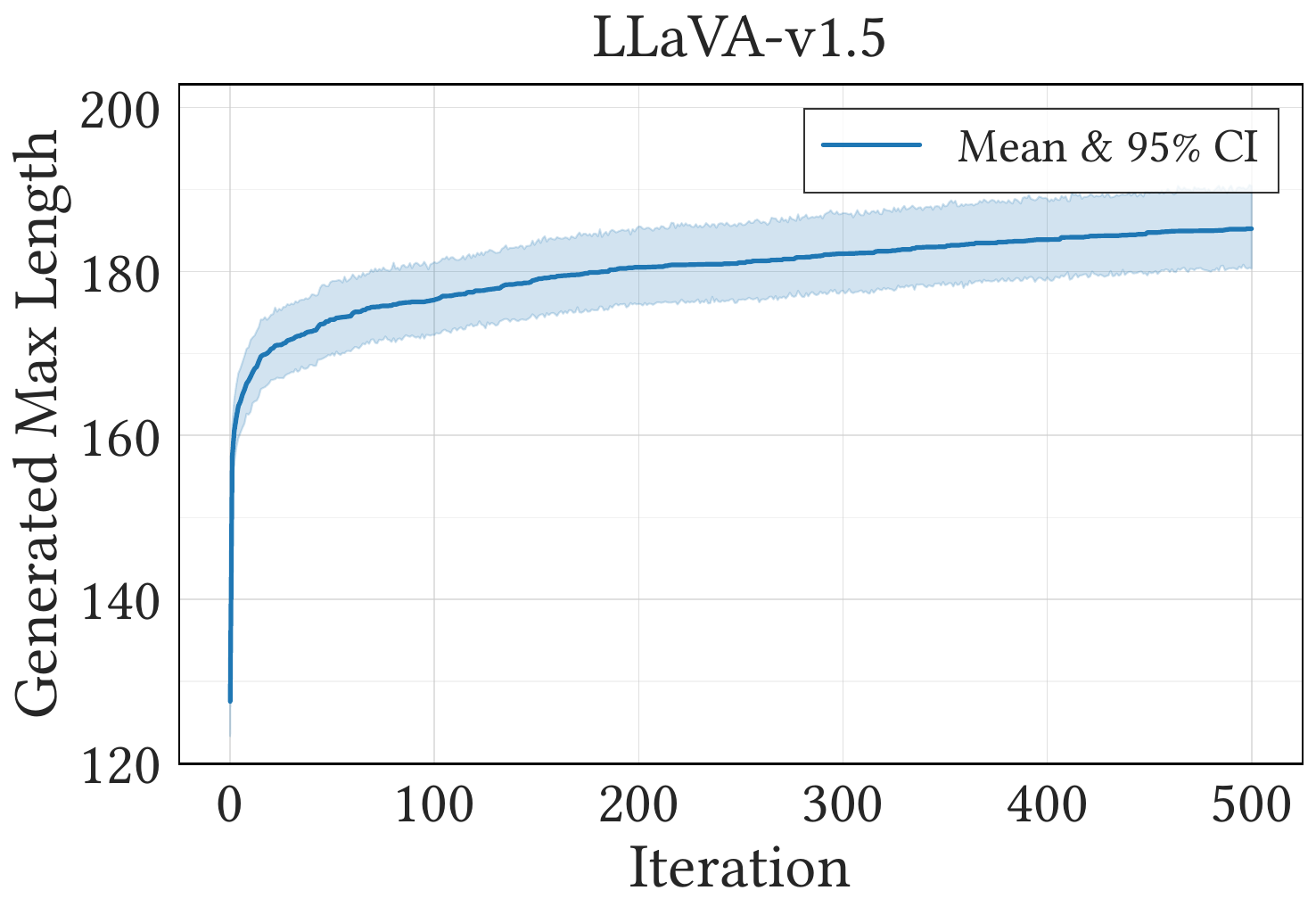}
\hfill
\includegraphics[width=0.32\textwidth]{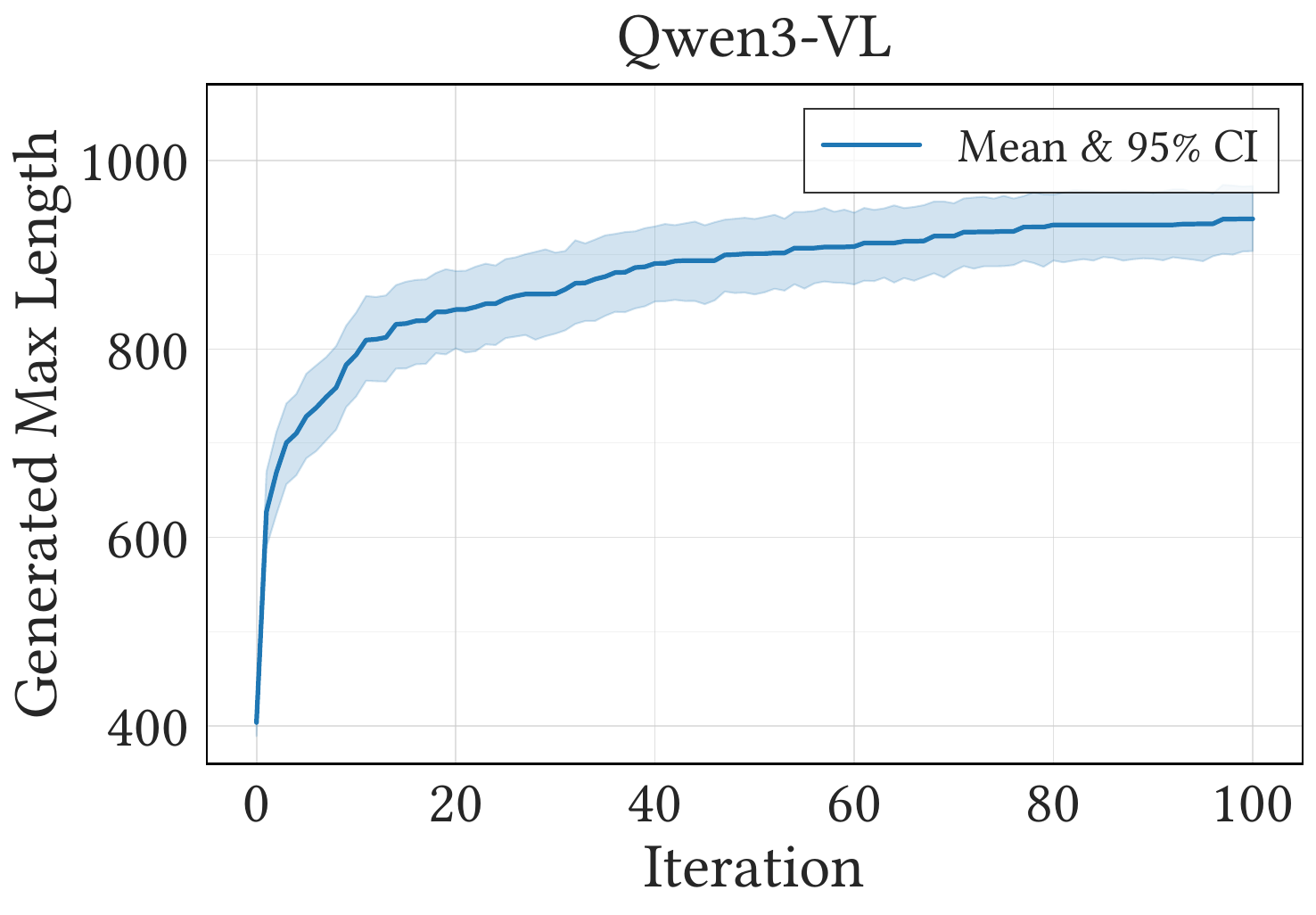}
\caption{Optimization trajectories of generated sequence length over Phase I iterations for BLIP, LLaVA, and Qwen models (mean $\pm$ 95\% CI). InternVL3-8B is omitted because most samples reach the maximum generation budget during the early optimization stage, resulting in a nearly saturated trajectory.}
\label{fig:progress}
\endgroup
\end{minipage}
\vspace{\baselineskip}
}]

\begin{chronoappendixfigure}
\centering
\includegraphics[width=\columnwidth]{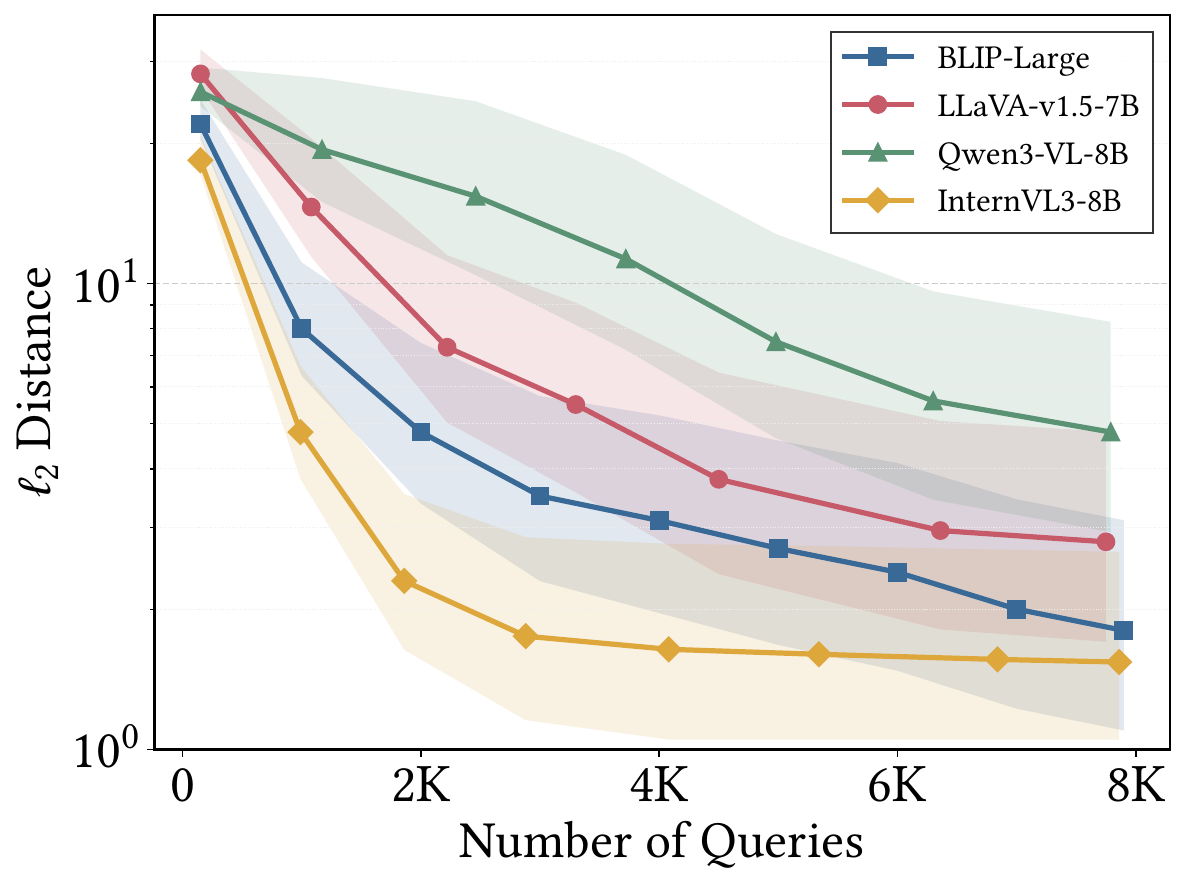}
\caption{Query efficiency comparison across different Vision-Language Models. The x-axis represents the number of model queries, and the y-axis shows the $\ell_2$ distance of the adversarial perturbation.}
\label{fig:query}
\end{chronoappendixfigure}

\mypara{Perturbation--Efficacy Relationship}
\autoref{fig:efficacy_scatter_comparison} jointly visualizes perturbation magnitude, generated response length, and inference time on Qwen3-VL-8B. This analysis complements the refinement trajectories by examining the relationship between perturbation distortion and attack effectiveness.

\begin{chronoappendixfigure}
\centering
\includegraphics[width=\columnwidth]{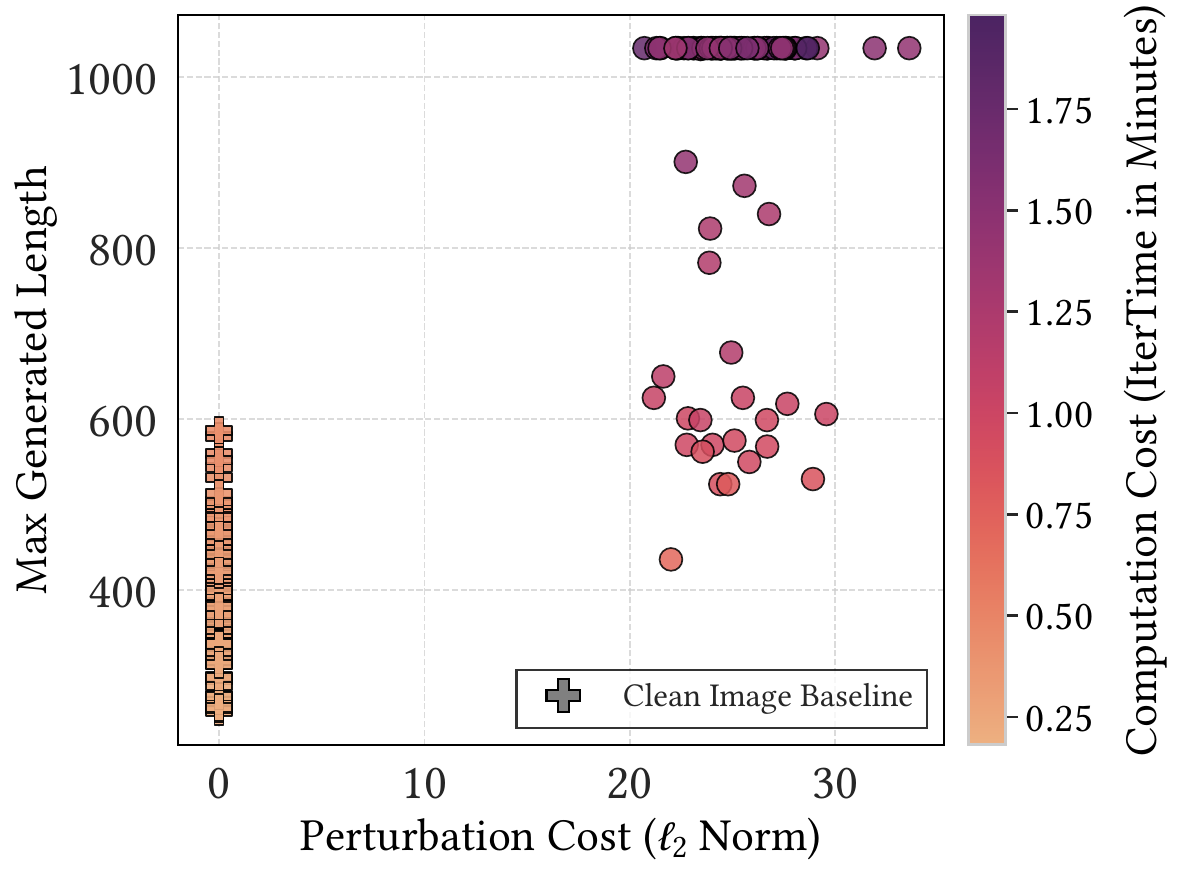}
\caption{Relationship between perturbation magnitude and attack effect on Qwen3-VL-8B. The x-axis denotes the masked $\ell_2$ perturbation norm, the y-axis denotes generated sequence length, and the color indicates per-iteration inference time.}
\label{fig:efficacy_scatter_comparison}
\end{chronoappendixfigure}

\mypara{Results on Flickr30K}
\autoref{tab:flickr_results} presents the full attack results on the Flickr30K dataset~\cite{plummer2015flickr30k}, complementing the MSCOCO and ImageNet results reported in the main paper. ChronoPatch maintains consistent effectiveness across all four target models, with $\Delta T$ ranging from 44.17\% on BLIP-Large to 291.33\% on InternVL3-8B. The results confirm that the attack's effectiveness is not dataset-specific and generalizes across diverse image distributions.

\begin{chronoappendixtable}
\centering
\caption{Attack results on Flickr30K across four target VLMs. $\Delta L$ and 
$\Delta T$ denote percentage increases over the clean baseline.}
\label{tab:flickr_results}
\resizebox{\columnwidth}{!}{
\begin{tabular}{@{}llcccc@{}}
\toprule
\textbf{Model} & \textbf{Method} & \textbf{Avg. Len} & \textbf{Len/Max} 
& \textbf{$\Delta L$ (\%)} & \textbf{$\Delta T$ (\%)} \\
\midrule
\multirow{2}{*}{BLIP-Large} 
  & Clean          & 22.51 & 4.40\% & ---       & ---       \\
  & ChronoPatch    & 32.06 & 6.26\% & 42.47\% & 44.17\% \\
\midrule
\multirow{2}{*}{LLaVA-v1.5-7B}
  & Clean          & 125.73 & 12.28\% & ---       & ---       \\
  & ChronoPatch    & 183.47 & 17.92\% & 45.92\% & 46.38\% \\
\midrule
\multirow{2}{*}{Qwen3-VL-8B}
  & Clean          & 407.21 & 39.77\% & ---       & ---       \\
  & ChronoPatch    & 921.34 & 89.97\% & 126.29\% & 138.51\% \\
\midrule
\multirow{2}{*}{InternVL3-8B}
  & Clean          & 243.18 & 23.75\% & ---       & ---       \\
  & ChronoPatch    & 1015.43 & 99.14\% & 317.56\% & 291.33\% \\
\bottomrule
\end{tabular}
}
\end{chronoappendixtable}

\mypara{Repetition Degeneration on Flickr30K}
\autoref{tab:repn_flickr} reports the $rep\text{-}n$ degeneration metrics on Flickr30K, consistent with the MSCOCO findings in the main paper. The pattern of severity across architectures is preserved: BLIP and LLaVA exhibit only marginal rises in repetition, while Qwen3-VL and InternVL3 undergo severe structural collapse, with InternVL3 reaching a $rep\text{-}4$ ratio of 58.27\% under 
adversarial conditions.

\begin{chronoappendixtable}
\centering
\caption{Text degeneration (\rep{n}) on Flickr30K under clean and adversarial 
settings.}
\label{tab:repn_flickr}
\resizebox{\columnwidth}{!}{
\begin{tabular}{@{}llcccc@{}}
\toprule
\textbf{Model} & \textbf{Setup} & \textbf{$\Delta L$} 
& \textbf{\rep{2}} & \textbf{\rep{3}} & \textbf{\rep{4}} \\
\midrule
\multirow{2}{*}{BLIP}      
  & Clean & \multirow{2}{*}{+42.47\%} & 0.19 & 0.05 & 0.01 \\
  & Adv   &                           & 4.51 & 2.91 & 2.03 \\
\midrule
\multirow{2}{*}{LLaVA}     
  & Clean & \multirow{2}{*}{+45.92\%} & 5.31 & 1.27 & 0.10 \\
  & Adv   &                           & 12.94 & 3.62 & 2.11 \\
\midrule
\multirow{2}{*}{Qwen3-VL}  
  & Clean & \multirow{2}{*}{+126.29\%} & 8.52 & 1.59 & 0.47 \\
  & Adv   &                            & 39.73 & 30.41 & 28.07 \\
\midrule
\multirow{2}{*}{InternVL3} 
  & Clean & \multirow{2}{*}{+291.33\%} & 14.91 & 12.87 & 9.94 \\
  & Adv   &                            & 69.43 & 62.11 & 58.27 \\
\bottomrule
\end{tabular}
}
\end{chronoappendixtable}

\mypara{Robustness to Input Transformations}
We evaluate four common input transformations on Qwen3-VL: JPEG compression, Gaussian blur, random resizing and padding, and feature quantization. As shown in \autoref{tab:defense_robustness}, all transformations attenuate the attack effect, reducing $\Delta L$ from 110.68\% to 45.48\%--67.88\% and $\Delta T$ from 142.42\% to 49.65\%--74.57\%. Nevertheless, substantial length and latency amplification persists under all tested transformations, indicating that standard input preprocessing alone is insufficient to eliminate the attack.

\begin{chronoappendixtable}
\centering
\caption{Robustness of ChronoPatch under common input transformations on Qwen3-VL. $\Delta L$ and $\Delta T$ denote the relative increases in generated length and end-to-end inference latency compared with clean-image inference.}
\label{tab:defense_robustness}
\small
\begin{tabular}{l c c}
\toprule
\textbf{Input Transformation} & \textbf{$\Delta L$ (\%)} & \textbf{$\Delta T$ (\%)} \\
\midrule
None & 110.68\% & 142.42\% \\
JPEG Compression  & 59.24\% & 67.13\% \\
Gaussian Blur & 56.03\% & 61.05\% \\
Random Resize \& Pad & 67.88\% & 74.57\% \\
Quantization & 45.48\% & 49.65\% \\
\bottomrule
\end{tabular}
\end{chronoappendixtable}

\mypara{Attribution Analysis}
We use Grad-CAM~\cite{selvaraju2017grad} on BLIP-Large to examine the visual activation patterns induced by ChronoPatch. As shown in \autoref{fig:grad_cam}, clean inputs produce activations concentrated on semantically relevant regions, whereas adversarial inputs exhibit shifted activations toward the localized perturbed regions. This observation provides qualitative evidence that delay anchors can influence the visual features emphasized during generation, although it does not establish a causal relationship.

\begin{chronoappendixfigure}
\centering
\includegraphics[width=\columnwidth]{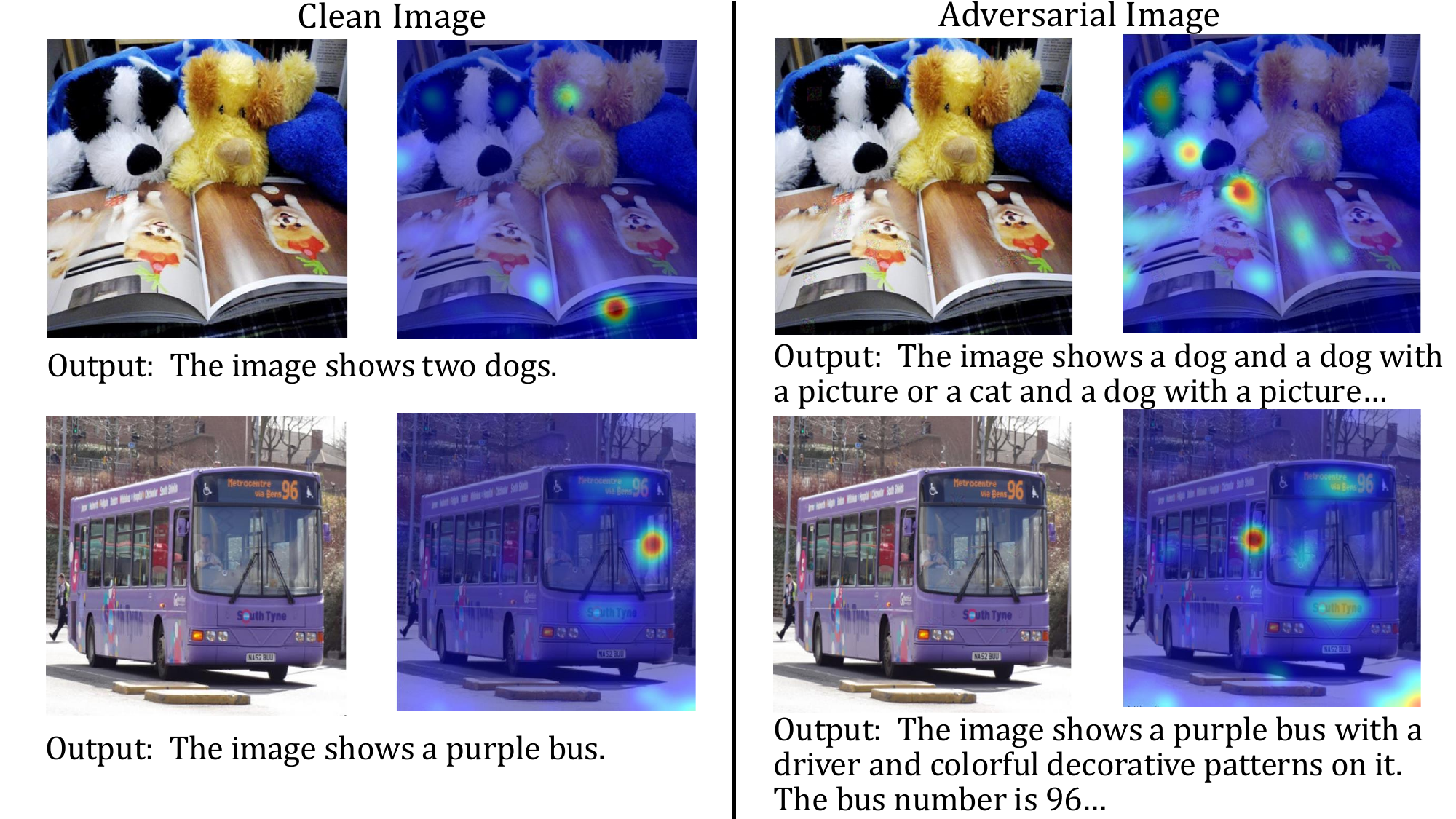}
\caption{Attribution-based supporting evidence using Grad-CAM on BLIP-Large. The heatmaps compare visual regions highlighted under clean and adversarial inputs. Adversarial images tend to shift part of the highlighted activation toward the localized perturbed region.}
\label{fig:grad_cam}
\end{chronoappendixfigure}

\clearpage
\section{Algorithmic Details}
\label{app:algorithms}

This section provides the complete procedures of the two-stage ChronoPatch pipeline. \autoref{alg:phase1_eogen_elitism} describes Phase~I delay-anchor discovery, which searches over patch textures and locations using final-response length as feedback and returns a visual delay anchor together with its spatial mask.
\autoref{alg:SC_hsja} presents Phase~II perturbation refinement, which starts from the discovered anchor and progressively reduces the perturbation magnitude while preserving the long-generation effect under the sample-specific response-length threshold.

\begin{ieeealgorithm}[htbp]{Phase I: Delay-Anchor Discovery}
\label{alg:phase1_eogen_elitism}
\begin{algorithmic}[1]
\REQUIRE Clean image $x_{\mathrm{clean}}$, generation budget $G$, candidate set size $S$, max iterations $T$, retained candidate count $K$
\ENSURE Visual delay anchor $x_{\mathrm{anchor}}$ and spatial mask $M_{\mathrm{anchor}}$

\STATE Initialize candidate set $P = \{\mathbf{g}^{(1)}, \dots, \mathbf{g}^{(S)}\}$ by uniform texture sampling and center-biased location sampling
\STATE Archived texture set $\mathcal{A} \gets \emptyset$, best reward $R_{\max} \gets -\infty$

\FOR{$t = 1$ \textbf{to} $T$}
    \FOR{\textbf{each} candidate $\mathbf{g}^{(i)} \in P$}
        \STATE Render perturbed image $\mathbf{I}^{(i)}$ and its mask $\mathbf{M}^{(i)}$ using $x_{\mathrm{clean}}$ and $\mathbf{g}^{(i)}$
        \STATE $L_{\mathrm{gen}} \gets \textsc{Query\_VLM\_API}(\mathbf{I}^{(i)}, G)$
        \STATE $R(\mathbf{I}^{(i)}) \gets L_{\mathrm{gen}}$
        \IF{$R(\mathbf{I}^{(i)}) > R_{\max}$}
            \STATE $R_{\max} \gets R(\mathbf{I}^{(i)})$
            \STATE $x_{\mathrm{anchor}} \gets \mathbf{I}^{(i)}$;
            $M_{\mathrm{anchor}} \gets \mathbf{M}^{(i)}$
            \STATE Update $\mathcal{A}$ with high-reward textures
        \ENDIF

        \IF{$R_{\max} \geq G$}
            \STATE \textbf{return} $x_{\mathrm{anchor}}, M_{\mathrm{anchor}}$
        \ENDIF
    \ENDFOR

    \STATE $P_{\mathrm{retain}} \gets \textsc{Select\_Top\_K}(P, K)$
    \STATE $\sigma_{\mathrm{pos}}(t), \sigma_{\mathrm{tex}}(t) \gets \textsc{Linear\_Anneal}(t, T)$
    \STATE $P_{\mathrm{next}} \gets P_{\mathrm{retain}}$

    \WHILE{$|P_{\mathrm{next}}| < S$}
        \STATE Generate a child candidate by crossover over two retained candidates
        \IF{$\textsc{Uniform}(0,1)<P_{\mathrm{inject}}$ \textbf{and} $\mathcal{A}\neq\emptyset$}
            \STATE Inject an archived high-reward texture from $\mathcal{A}$ into the child candidate
        \ELSE
            \STATE Mutate positions and textures using $\sigma_{\mathrm{pos}}(t)$ and $\sigma_{\mathrm{tex}}(t)$
        \ENDIF
        \STATE Add the child candidate to $P_{\mathrm{next}}, M_{\mathrm{anchor}}$
    \ENDWHILE
    \STATE $P \gets P_{\mathrm{next}}$
\ENDFOR

\STATE \textbf{return} $x_{\mathrm{anchor}}, M_{\mathrm{anchor}}$
\end{algorithmic}
\end{ieeealgorithm}

\begin{ieeealgorithm}[htbp]{Phase II: Perturbation Refinement}
\label{alg:SC_hsja}
\begin{algorithmic}[1]
\REQUIRE Clean image $x_{\mathrm{clean}}$, visual delay anchor $x_{\mathrm{anchor}}$, mask $M$, threshold $\tau$, target VLM $\mathcal{F}$
\ENSURE Refined adversarial image $x_{\mathrm{adv}}$

\STATE $x_t \gets x_{\mathrm{anchor}}$
\STATE $x_{\mathrm{best}} \gets x_{\mathrm{anchor}}$
\STATE $d_{\mathrm{eff}} \gets 3 \cdot \sum(M)$

\FOR{$t = 1$ \textbf{to} $T_{\max}$}
     \STATE $x_{\mathrm{bd}} \gets \textsc{Binary\_Search}(x_t, x_{\mathrm{clean}}, M, \tau, \mathcal{F})$
    
    \STATE $\delta_t \gets \textsc{Compute\_Base\_Delta}(x_{\mathrm{bd}}, x_{\mathrm{clean}})$
    \STATE $\delta_t^* \gets \delta_t \cdot \sqrt{D / d_{\mathrm{eff}}}$
    
    \STATE $\mathit{retries} \gets 0$
    \WHILE{$\mathit{retries} < \mathit{max\_retries}$}
        \STATE $U \gets \textsc{Generate\_Masked\_Noise}(B, M)$
        \STATE $\Phi \gets \textsc{Query\_VLM}(x_{\mathrm{bd}} + \delta_t^* U, \tau)$
        \IF{$\text{mean}(\Phi) \leq -0.95$}
            \STATE $\delta_t^* \gets \min(\delta_t^* \cdot 3, \delta_{\max})$
            \STATE \textbf{continue}
        \ELSIF{$\text{mean}(\Phi) \geq 0.95$}
            \STATE $\delta_t^* \gets \max(\delta_t^*/2, \delta_{\min})$
        \ELSE
            \STATE \textbf{break}
        \ENDIF
        \STATE $\mathit{retries} \gets \mathit{retries}+1$
    \ENDWHILE
    
    \STATE $g_t \gets \textsc{Compute\_Baseline\_Gradient}(\Phi, U)$
    
    \STATE $d_t \gets \textsc{Select\_Direction}(g_t, -g_t, x_{\mathrm{bd}}, x_{\mathrm{clean}})$
    
    \STATE $\xi \gets \|x_{\mathrm{bd}}-x_{\mathrm{clean}}\|_2/\sqrt{t}$
    \STATE $\mathit{fail\_count} \gets 0$
    \WHILE{$\textsc{Query\_VLM}(x_{\mathrm{bd}}+\xi d_t,\tau) == -1$}
        \STATE $\xi \gets \xi/2$
        \STATE $\mathit{fail\_count} \gets \mathit{fail\_count}+1$
        \IF{$\mathit{fail\_count}>\mathit{Max\_Fails}$}
            \STATE \textbf{break}
        \ENDIF
    \ENDWHILE
    
    \STATE $x_{t+1} \gets \textsc{Clip}(x_{\mathrm{bd}}+\xi d_t,M)$
    
    \IF{$\phi(x_{t+1})=+1$}
        \IF{$\|x_{t+1}-x_{\mathrm{clean}}\|_2
        < \|x_{\mathrm{best}}-x_{\mathrm{clean}}\|_2$}
            \STATE $x_{\mathrm{best}} \gets x_{t+1}$
        \ENDIF
    \ENDIF
    
    \STATE $x_t \gets x_{t+1}$
\ENDFOR

\STATE \textbf{return} $x_{\mathrm{best}}$
\end{algorithmic}
\end{ieeealgorithm}

\end{document}